\documentclass[review]{elsarticle}
\usepackage{geometry}
\usepackage{lineno,hyperref}
\usepackage{supertabular} 
\usepackage{graphicx}
\usepackage{amsmath}
\usepackage{amsfonts}
\usepackage{amssymb}
\journal{Elsevier}

\begin{document}

\begin{frontmatter}

\title{Authentication Protocols for Internet of Things: A Comprehensive Survey}

\author[adress1]{Mohamed Amine Ferrag\corref{mycorrespondingauthor}}
\cortext[mycorrespondingauthor]{Corresponding author}
\ead{mohamed.amine.ferrag@gmail.com}

\author[adress2]{Leandros A. Maglaras}
\ead{Leandros.maglaras@dmu.ac.uk}

\author[adress2]{Helge Janicke}
\ead{heljanic@dmu.ac.uk}

\author[adress3]{Jianmin Jiang}
\ead{jianmin.jiang@szu.edu.cn}

\address[adress1]{Department of Computer Science, Guelma University, BP 401, 24000, Algeria}
\address[adress2]{Cyber Security Centre, School of Computer Science and Informatics,  De Montfort University, Leicester, United Kingdom}
\address[adress3]{Research Institute for Future Media Computing, Shenzhen University, Shenzhen, China}

\begin{abstract}
In this paper, we present a comprehensive survey of authentication protocols for Internet of Things (IoT). Specifically, we select and in-detail examine more than forty authentication protocols developed for or applied in the context of the IoT under four environments, including: (1) Machine to machine communications (M2M), (2) Internet of Vehicles (IoV), (3) Internet of Energy (IoE), and (4) Internet of Sensors (IoS). We start by reviewing all survey articles published in the recent years that focusing on different aspects of the IoT idea. Then, we review threat models, countermeasures, and formal security verification techniques used in authentication protocols for the IoT. In addition, we provide a taxonomy and comparison of authentication protocols for the IoT in form of tables in five terms, namely, network model, goals, main processes, computation complexity, and communication overhead. Based on the current survey, we identify open issues and suggest hints for future research.
\end{abstract}


\end{frontmatter}

%\linenumbers

\section{Introduction}
Researchers lately refer to the term "The Internet of Things (IoT)" as the revolution of the future. According to forecasts from Cisco Systems \cite{283}, by 2020, the Internet will consist of over 50 billion connected things, including, sensors, actuators, GPS devices, mobile devices, and all smart things that we can envision in the future. Currently, IBM has decided to combine several products and services into an offer called IoT Solutions Practice \cite{284} to allow the customers to find all IBM IoT offers at the same location. For example, IBM offers the Watson IoT platform \cite{285}, which combines scanning, security, and blockchain technology for authentication with a set of APIs such as IBM's SoftLayer cloud infrastructure \cite{286}. The IoT can be realized under three scopes, namely, internet-oriented (middleware), things-oriented (sensors) and semantic-oriented (knowledge) \cite{19}. According to Atzori et al. \cite{2}, IoT can be represented as a three-layered architectural model, which consists of the application layer, the network layer, and the sensing layer.

\begin{figure}[h]
 \centering
 \includegraphics[width=0.6\linewidth]{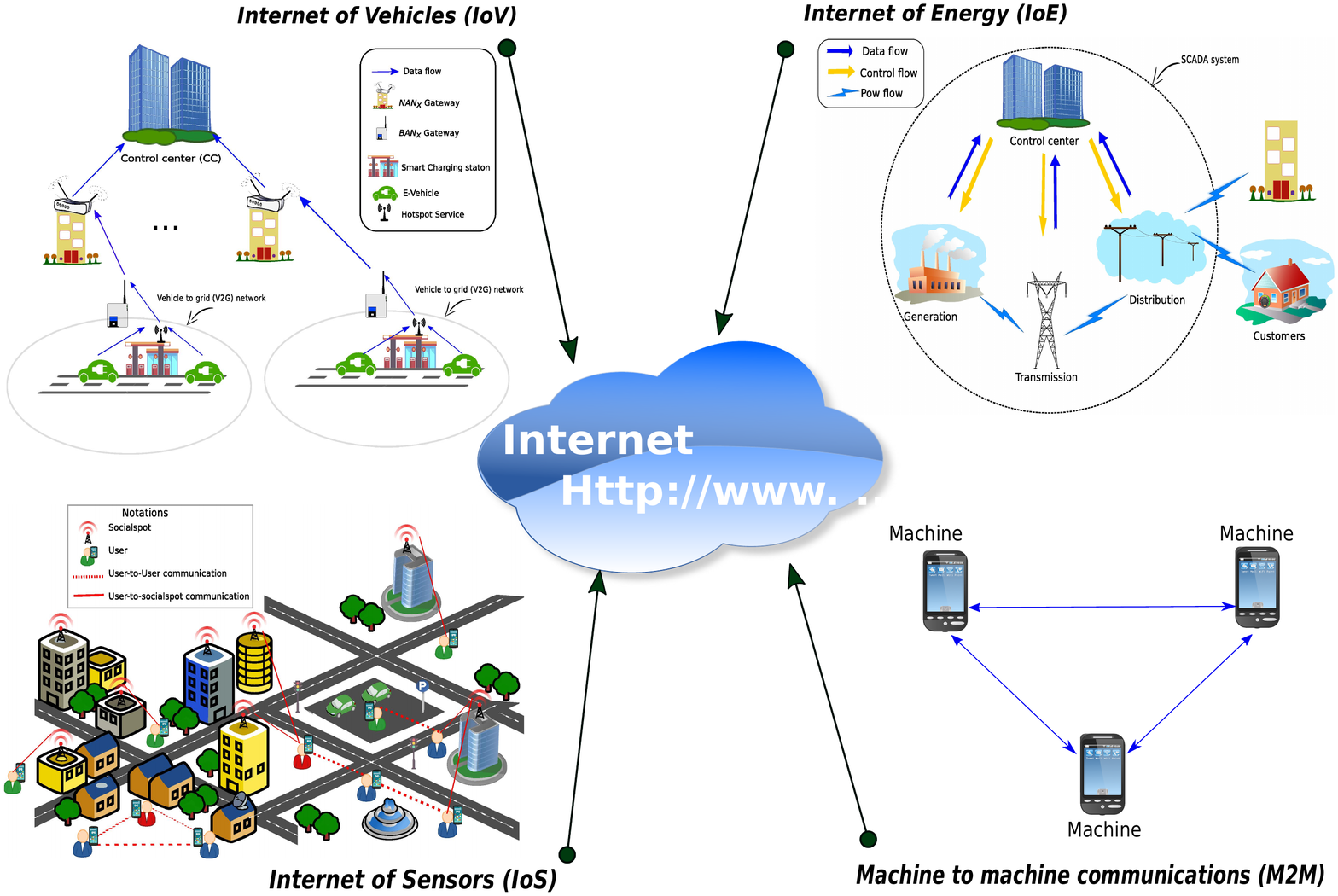}
 \caption{Internet of Things (IoT) in four environments, including: (1) Machine to machine communications (M2M), (2) Internet of Vehicles (IoV), (3) Internet of Energy (IoE), and (4) Internet of Sensors (IoS)}
 \label{fig:Fig1a}
 \end{figure}
\begin{figure}[h]
 \centering
 \includegraphics[width=0.6\linewidth]{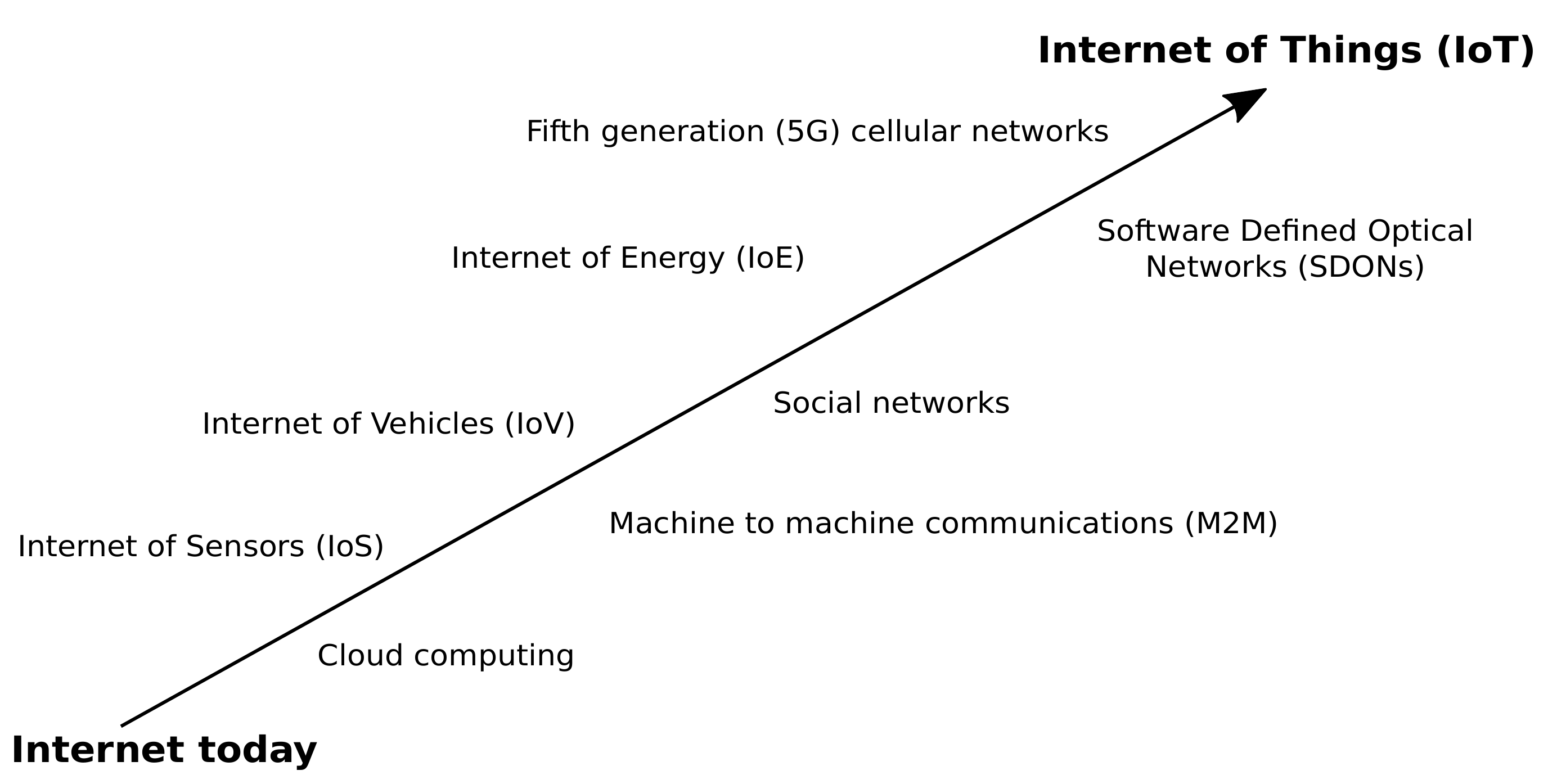}
 \caption{Vision of the IoT with main features and challenges}
 \label{fig:Fig1b}
 \end{figure}

As shown in Fig. \ref{fig:Fig1a}, IoT has made its entrance in four fields, including, (1) Machine to machine communications (M2M), (2) Internet of Vehicles (IoV), (3) Internet of Energy (IoE), and (4) Internet of Sensors (IoS). M2M is a technology crucial for the realization of IoT, which is based on different protocols such as the protocol Stack \cite{288}. The IoV is based on the concept of Vehicular Cloud, which offers access to the Internet, and is temporarily created by inter-connecting resources available on the vehicles along with Road Side Units (RSUs) \cite{287,292,293}. According to ARTEMIS-project \cite{289}, the IoE is the connection of smart grids with the internet in order to enable intelligent control of energy production, storage and distribution. The IoS refers to connecting sensors with the Internet using ZigBee and other IEEE 802.15.4 based protocols \cite{290}.

The vision of the IoT will advance based on many new features and coping with novel challenges, as shown in Fig. \ref{fig:Fig1b}, including, cloud computing, M2M, IoS, IoE, IoV, social networks, software defined optical networks (SDONs), and fifth generation (5G) cellular networks. The IoT data which will be produced from billions of interactions between devices and people is not only going to be massive, but also complex and it will suffer from many security and privacy problems, especially regarding the authentication between devices. To resolve these security issues, researchers in the field of computer security have developed many authentication protocols for or applied in the context of the IoT. In a recent survey paper which was published in 2015 \cite{45} authors reviewed the state of the art of RFID authentication schemes which used elliptic curve cryptography and were used in healthcare environments. The aim of our survey paper is to provide a comprehensive and systematic review of recent studies on published authentication protocols for the IoT in four environments, including, M2M, IoV, IoE, and IoS. More precisely, we select and in-detail examine more than forty authentication protocols. The original set of papers was formed from the searchers run on SCOPUS and Web of Science from the period between 2005 and 2016 (See Tab. \ref{tab:Tab1a} for a breakdown of publication dates). The main contributions of this paper are:

\begin{itemize}
\item  We survey around seventy survey articles published in the recent years that deal with the IoT.
\item  We discuss the authentication protocols in M2M, IoV, IoE, and IoS that were evaluated under thirty-five attacks. Then, we focus on four attacks, which are mostly studied in earlier works, namely, man-in-the-middle attack, impersonation attack, forging attack, and replay attack.
\item  We present various countermeasures and formal security verification techniques used by authentication protocols for the IoT.
\item  We present a side-by-side comparison in a tabular form of the current state-of-the-art of authentication protocols (more than forty) proposed for the IoT from five different aspects, namely, network model, goals, main processes, computation complexity, and communication overhead.
\item  We discuss the open issues for M2M, IoV, IoE, and IoS. 
\end{itemize}

The remainder of this paper is organized as follows. In section \ref{sec:surveys-articles-for-the-iot}, we summarize the existing survey works on different aspects of the IoT idea. In section \ref{sec:threat-models}, we present an overview of threat models in the IoT. Then in section \ref{sec:countermeasures-and-formal-security-verification-techniques}, we discuss the countermeasures and formal security verification techniques. In section \ref{sec:taxonomy-and-comparison-of-authentication-protocols-for-the-iot}, we present a side-by-side comparison in a tabular form for the current state-of-the-art of authentication protocols proposed for M2M, IoV, IoE, and IoS. Finally, we identify the future directions \ref{sec:open-issues} and conclude the article \ref{sec:conclusion}.

\begin{table}
\centering
\caption{Publication date breakdown - Surveyed papers (authentication protocols)}
{\tiny
\begin{tabular}{|p{3.5in}|p{0.3in}|} \hline 
Papers & Year  \\ \hline 
\cite{249} & 2005 \\ \hline 
\cite{256}\cite{259} & 2006 \\ \hline 
\cite{255} & 2007 \\ \hline 
\cite{254} \cite{247} & 2008 \\ \hline 
\cite{121} \cite{258} \cite{216} \cite{268} \cite{215} \cite{269} \cite{270} \cite{267} & 2010 \\ \hline 
\cite{250} \cite{122} \cite{260} \cite{127} \cite{252} \cite{140} \cite{206} & 2011 \\ \hline 
\cite{77} \cite{251} \cite{81} \cite{166} \cite{204} & 2012 \\ \hline 
\cite{136} \cite{79} \cite{89} \cite{90} \cite{203} \cite{195} \cite{263} \cite{264} \cite{205} & 2013 \\ \hline 
\cite{78} \cite{118} \cite{95} \cite{128} \cite{213} \cite{202} & 2014 \\ \hline 
\cite{108} \cite{117} \cite{125} \cite{134} \cite{262} \cite{196} \cite{265} \cite{266} \cite{197} & 2015 \\ \hline 
\cite{76} \cite{71} \cite{91} \cite{92} \cite{93} \cite{94} \cite{135} \cite{167} \cite{168} \cite{170} \cite{171} \cite{172} \cite{173} \cite{174} \cite{175} \cite{177} \cite{179} & 2016 \\ \hline 
\cite{176} & 2017 \\ \hline 
\end{tabular}
}
\label{tab:Tab1a}
\end{table}

\section{Surveys articles for the IoT}\label{sec:surveys-articles-for-the-iot}
There exist around seventy survey articles published in the recent years that deal with Internet of Things, focusing on different aspects of the IoT idea, e.g. networking, applications, standardization, social interactions, security and many more. These survey articles are categorized in terms of field of research and year of publication as shown in tables \ref{tab:Tab2a} and \ref{tab:Tab2b}. As it is seen from Tab. \ref{tab:Tab2b} the Internet of Things concepts attracts more and more attention as the years pass by and although a lot of different areas related to IoT are covered from previous review works, no survey article exists that thoroughly investigates authentication protocols that are especially developed for this new technology or better say this blend of technologies and systems. In this section we will briefly present all these survey articles grouped as shown in Tab. \ref{tab:Tab2a} and will discuss in more depth previous work that deal with security and privacy issues of the IoT.

\begin{table}[!]
\centering
\caption{Areas of research of each survey article for the IoT}
{\tiny
\textbf{DD}: Data quality and Database Management, \textbf{MW}: Middleware, \textbf{AP}: Applications, \textbf{HD}: Healthcare domain, \textbf{SE}: Smart environments, \textbf{SP}: Security and privacy, \textbf{Exp}: Experimentation, \textbf{Net}: Networking, \textbf{ST}: Standardization, \textbf{Arch}: Architecture \textbf{SR}: Searching, \textbf{RFID}: RFID technology, \textbf{Soc}: Social internet of things, \textbf{CoA}: Context-awareness, \textbf{DM}: Data mining, \textbf{IIoT}: Industrial Internet of Things
 \begin{tabular}{|p{2in}|p{0.1in}|p{0.15in}|p{0.15in}|p{0.15in}|p{0.15in}|p{0.15in}|p{0.15in}|p{0.15in}|p{0.15in}|p{0.15in}|p{0.15in}|p{0.2in}|p{0.15in}|p{0.15in}|p{0.15in}|p{0.15in}|} \hline 
\textbf{Ref.} & \textbf{DD} & \textbf{MW} & \textbf{AP} & \textbf{HD} & \textbf{SE} & \textbf{SP} & \textbf{Exp} & \textbf{Net} & \textbf{ST} & \textbf{Arch} & \textbf{SR} & \textbf{RFID} & \textbf{Soc} & \textbf{CoA} & \textbf{DM} & \textbf{IIoT} \\ \hline 
\cite{1,61,65,68} & \checkmark &  &  &  &  &  &  &  &  &  &  &  &  &  &  &  \\ \hline 
\cite{2,7,10,12,55} &  & \checkmark &  &  &  &  &  &  &  &  &  &  &  &  &  &  \\ \hline 
\cite{2,48,64}  &  &  &  & \checkmark &  &  &  &  &  &  &  &  &  &  &  &  \\ \hline 
\cite{2,33,57}  &  &  &  &  & \checkmark &  &  &  &  &  &  &  &  &  &  &  \\ \hline 
\cite{5}  &  &  &  &  &  &  & \checkmark &  &  &  &  &  &  &  &  &  \\ \hline 
\cite{6,13,20,53,66,67,68} &  &  &  &  &  &  &  & \checkmark &  &  &  &  &  &  &  &  \\ \hline 
\cite{8,9,18,21,26,37} &  &  &  &  &  &  &  &  & \checkmark &  &  &  &  &  &  &  \\ \hline 
\cite{11,28,31,34,62}  &  &  &  &  &  &  &  &  &  & \checkmark &  &  &  &  &  &  \\ \hline 
\cite{16}  &  &  &  &  &  &  &  &  &  &  & \checkmark &  &  &  &  &  \\ \hline 
\cite{3,4,9,13,14,20,24,29,32,35,39,40,41,45,47,49,50,56,63,70} &  &  &  &  &  & \checkmark &  &  &  &  &  &  &  &  &  &  \\ \hline 
\cite{17,52}  &  &  &  &  &  &  &  &  &  &  &  & \checkmark &  &  &  &  \\ \hline 
\cite{23,25,42,69}  &  &  &  &  &  &  &  &  &  &  &  &  & \checkmark &  &  &  \\ \hline 
\cite{2,8,9,11,15,19,20,22,36,38,43,44,46,54,58,59,60} &  &  & \checkmark &  &  &  &  &  &  &  &  &  &  &  &  &  \\ \hline 
\cite{27}  &  &  &  &  &  &  &  &  &  &  &  &  &  & \checkmark &  &  \\ \hline 
\cite{30,51}  &  &  &  &  &  &  &  &  &  &  &  &  &  &  & \checkmark &  \\ \hline 
\cite{28} &  &  &  &  &  &  &  &  &  &  &  &  &  &  &  & \checkmark \\ \hline 
\end{tabular}
}
\label{tab:Tab2a}
\end{table}

\begin{table}
\centering
\caption{Year of publication}
{\tiny \begin{tabular}{|p{3.7in}|p{0.4in}|} \hline 
\textbf{Ref.} & \textbf{Year} \\ \hline 
\cite{1}\textbf{} & 2009 \\ \hline 
\cite{2,3,4} & 2010 \\ \hline 
\cite{5,6,7,8,12,13,15,16} & 2011 \\ \hline 
\cite{9,10,11,14,17,22,23} & 2012 \\ \hline 
\cite{18,19,20,21,24,26} & 2013 \\ \hline 
\cite{27,28,29,30,31,32,33,34,35,36,37,38,39,41,42,43} & 2014 \\ \hline 
 \cite{44,45,46,47,48,49,50,51,52,53} & 2015 \\ \hline 
\cite{40,54,55,56,57,58,59,60,61,62,63,64,65,66,67,68,69,70} & 2016 \\ \hline 
\end{tabular}}
\label{tab:Tab2b}
\end{table}

\begin{table}[!]
\centering
\caption{A comparison of related surveys in the literature (Surveys on Security and Privacy for the IoT)}
{\tiny \checkmark :indicates fully supported; X: indicates not supported; 0: indicates partially supported. 
\begin{tabular}{|p{1.5in}|p{0.7in}|p{0.7in}|p{4in}|} \hline 
\textbf{Survey on Security and Privacy for the IoT} & \textbf{Privacy preserving schemes} & \textbf{Authentication Protocols} & \textbf{Comments} \\ \hline 
Weber (2010) \cite{3} & 0 & X & - Presented milestones of an adequate legal framework for IoT privacy. \\ \hline 
Medaglia and Serbanati (2010) \cite{4} & 0 & X & - Presented a Short-Term and Long-Term vision for IoT privacy. \\ \hline 
Roman et al. (2011) \cite{13}  & X & X & - Analyzed some key management systems for sensor networks in the context of the IoT (public key cryptography and pre-shared keys). \\ \hline 
Miorandi et al. (2012) \cite{9} & 0 & X & - Presented some security challenges in IoT, including, Data confidentiality, Privacy, and Trust. \\ \hline 
Suo et al. (2012) \cite{14} & X & X & - Discussed the security requirements in each level for IoT (four key levels, i.e., recognition layer, network layer, support layer, and application layer) \\ \hline 
Aggarwal et al. (2013) \cite{20} & 0 & X & - Discussed the privacy in data collection, and during data transmission and sharing. \\ \hline 
Roman et al. (2013) \cite{24} & X & X & - Presented the security issues in distributed IoT systems. \\ \hline 
Yan et al. (2014) \cite{29} & \checkmark & X & - Surveyed the privacy-preserving schemes IoT, including, database query, scientific computations, intrusion detection, and data mining. \\ \hline 
Jing et al. (2014) \cite{32} & X & X & - Discussed the security issues and technical solutions in WSNs. \\ \hline 
Chabridon et al. (2014) \cite{35} & \checkmark & X & - Surveyed the state of the art of privacy technology from the perspective of the IoT. \\ \hline 
Ziegeldorf et al. \cite{39} & \checkmark & X & - Surveyed the privacy threats and challenges in the IoT. \\ \hline 
Keoh et al. (2014) \cite{41} & X & X & - Presented an overview of the efforts in the IETF to standardize security solutions for the IoT ecosystem. \\ \hline 
Sicari et al. (2015) \cite{45} & 0 & X & - Discussed the privacy, trust, enforcement, secure middleware, and mobile security in the IoT \\ \hline 
Granja et al. (2015) \cite{47} & X & 0 & - Discussed IoT communications and security at the physical and MAC layers. \\ \hline 
Sadeghi et al. (2015) \cite{49} & X & X & - Discussed an introduction to Industrial IoT systems with the related security and privacy challenges. \\ \hline 
Nguyen et al. (2015) \cite{50}  & 0 & X & - Surveyed the secure communication protocols for the IoT, including, asymmetric key schemes and symmetric key pre-distribution schemes. \\ \hline 
He et al. (2015)\cite{52}  & X & 0 & - Analyzed only the RFID authentication schemes for the IoT in healthcare environment using elliptic curve cryptography \\ \hline 
Xie et al. (2016) \cite{40} & X & X & - Reviewed the security issues for Web of Things. \\ \hline 
Singh et al. (2016) \cite{56} & X & X & - Analyzed the state of cloud-supported IoT to make explicit the security considerations. \\ \hline 
Li et al. (2016) \cite{63} & X & X & - Analyzed the security requirements and potential threats in a four-layer architecture for the IoT \\ \hline 
Airehrour et al. (2016)  \cite{70} & X & X & - Analyzed the security of routing protocols for the IoT. \\ \hline 
Our Work  &  0 & \checkmark & - Surveyed the authentication protocols for the IoT in four environments, including: (1) Machine to machine communications (M2M), (2) Internet of Vehicles (IoV), (3) Internet of Energy (IoE), and (4) Internet of Sensors (IoS). \\ \hline 
\end{tabular}}
\label{tab:Tab2c}
\end{table}

The first survey article in the literature that was dealing with the IoT concept was published back in 2009 by Cooper et al. \cite{1} and focused on the challenges for database management in the IoT. Seeing the IoT from that point of view they found that the technical priorities that needed to be addressed in order to support the interconnection of every device was proper indexing, archiving, development of smart agents the use of XML for achieving Interoperability and novel systems that will be able to offer efficient and secure transaction management. In a later survey article that was published in 2010, Atzori et al. \cite{2}, discussed the vision of `anytime, anywhere, any media, anything' communications that the IoT would bring in our everyday lives. Based on their research they had spotted two important technologies that needed to be applied in order to bring IoT into life, Internet Protocol version 6 (IPv6) and Web 2.0.  The same year, the first survey article that dealt with security and privacy issues related to IoT was published \cite{3}. In this article Weber et al., discussed the different measures that were needed in order to ensure the architecture's resilience to attacks, data authentication, access control and client privacy. The article dealt with security and privacy issues from the legislation perspective mostly due to the fact that the IoT was more an idea back in 2010 that a concrete system yet. Another article dealing with security and privacy was published in 2010 from Medaglia et al. \cite{4}. The article tried to present a short term and a long term vision of the IoT along with the security issues and solutions that would be needed.

In 2011 there were published eight survey articles that focused on the IoT \cite{5,6,7,8,12,13,15,16}. In \cite{5} authors conducted a thorough analysis of the different publicly available testbeds. In \cite{6} Mainetti et al., discussed about the necessary standards and solutions needed to guarantee the integration among several heterogeneous WSNs thus enabling smart objects to participate to the IoT. Bandyopadhyay et al. in \cite{7}, surveyed the state-of-the-art of the different approaches that middleware solutions used in order to support some of the functionalities necessary for operate in the IoT domain. Similar to this work, authors in \cite{12} surveyed the major challenges posed to service-oriented middleware towards sustaining a service-based Internet of Things.  Bandyopadhyay et al. \cite{8}, published an interesting survey article about the current developments related to IoT  and the open issues back in 2011. The article managed to spot most of the challenges that IoT had and still has to face nowadays, e.g. managing large amount of information and mining large volume of data, managing heterogeneity, ensuring security privacy and trust among others.  Feasible solutions for the problem of establishing a session key between a client and a server in the context of the Internet of Things were surveyed in \cite{13}, where the authors considered the scenario where at least one peer were sensor nodes. They especially focused on different cryptography solutions and how these could be applied to server and client nodes.  Ma et al. in \cite{15} gave an overview of the objectives of the IoT and the challenges involved in IoT development while in \cite{16} Zhang et al., covered the topic of how to build an appropriate search engine for IoT, a topic that was spotted from Cooper in \cite{1} back in 2009 as a challenge to be addressed in the future.

During 2012 and 2013 the following fourteen survey articles were published \cite{9,10,11,14,17,18,19,20,21,22,23,24,25,26} dealing with standardization, applications, architecture, security and privacy issues of the IoT. Articles \cite{18,21,26} surveyed standardization issues and how the IETF Constrained RESTful Environments (CoRE) working group focuses on facilitating the integration of constrained devices with the Internet at the service level. These articles pointed out that all the standardized protocols are only a starting point for exploring additional open issues such as resource representation, security and privacy, energy efficiency and so on. Authors in \cite{9,19} gave a general overview of the current vision, applications, architectural elements and future challenges and directions of the IoT. Miorandi et al. in \cite{9} discussed the potential impact of the IoT on smart home automation, smart cities, environmental monitoring, health care, smart businesses and security and surveillance making very clear, maybe for the first time, that the IoT concept involves every current or future technology that is going to be introduced in order to make our life better.  Authors in \cite{10} published a survey about middleware solutions for IoT similar to \cite{7,12}, presenting all the technical challenges of designing middleware systems for the IoT. Domingo et al. in \cite{11} performed a more narrow but extensive survey of the IoT for people with disabilities. Authors spotted the relevant application scenarios and main benefits along with the key research challenges, like customization, self management and security and privacy issues. They argued that as brain--computer interfaces (BCIs) are becoming commercial, they will also be a part of the IoT world. Articles \cite{14,24} focused on security and privacy issues as they were identified back in 2012 and 2013 respectively. Both articles agree that key management, legislation while authors in \cite{24} take one step further and propose that grouping of the IoT devices and creating the so called intranet of things could help impose security mechanisms more effectively. Survey articles \cite{17,20,22,23,25} researched the IoT system from different perspectives. In \cite{17} authors discussed about RFID technology and its applications in the IoT and in \cite{22} the role of fog computing in the IoT.  Aggarwal et al. \cite{20}, conducted a thorough research about data analytics, data mining and data management in the IoT world.  The article discusses all the issues related to data analytics and IoT, like real time and big data analytics challenges, effective crawling and searching in massive databases, privacy issues in data collections, data sharing and management and data security issues. Finally articles \cite{23} and \cite{25} survey for the first time the social concept of the IoT, the so called Social Internet of Things, a concept that later will raise a lot of attraction and research works.

During 2014 and 2015 more than twenty five new survey articles about IoT were published \cite{27,28,29,30,31,32,33,34,35,36,37,38,39,41,42,43,44,45,46,47,48,49,50,51,52,53}. Except from articles that discussed general issues regarding IoT \cite{36,37,43,44},  e.g. applications, challenges, trends and open issues, other papers focused on specific applications or research areas that are connected to the IoT idea. Authors in \cite{27,31,35,46} surveyed context awareness from the IoT perspective covering several aspects, like context-aware computing \cite{27}, context-aware product development \cite{31} and quality of context and privacy \cite{35}. Authors in all three articles agree that IoT thus brings new opportunities by enabling enriched context-aware services, but it also raises new challenges that need to be addressed. Zanella et al. \cite{33} focused specifically to an urban IoT system which is another term to describe the smart city environment. In contrast to the previous years during 2014 and 2015 a big proportion of the survey articles focus on security and privacy issues related to the IoT \cite{29,32,39,41,45,47,49,50},  revealing the significance that security was beginning to have for Cyber Physical systems.  Cyber Physical systems need to rely on IoT enabled technologies which can be effectively and efficiently supported and assisted by Cloud Computing infrastructures and platforms. The integration of IoT and Cloud computing was thoroughly surveyed from Botta et al. \cite{38} where also the possibility of exploiting fog computing capabilities for supporting the IoT concept was discussed.  Data mining in the IoT context were surveyed by Tsai et al. \cite{30} and Chen et al. \cite{51}.  Authors in \cite{30} presented a good summary of the potentials of applying data mining technologies to the IoT could have to people, the system itself, and other interconnected systems. Authors in \cite{51} took a step further and based on their survey and analysis proposed a big data mining system for IoT. Ortiz et al. \cite{42}, surveyed the Social Internet of Things and compared to the earlier survey articles \cite{23,25} proposed a generic SIoT architecture which consists of actors, a central intelligent system, an interface and the internet.  Two articles focused on IoT-based health care technologies \cite{52,48}, covering new platforms, applications and security and privacy issues that arise. Authors in \cite{28} conducted an extensive literature review about the current status and future research opportunities regarding the use of IoT in industries, the so called Industrial Internet of Things (IIoT) while in \cite{34} authors tried to identify the impact of the Internet of Things (IoT) on Enterprise Systems in modern manufacturing. 

During 2016 over fifteen new survey articles that focused on the IoT concept were published \cite{40,54,55,56,57,58,59,60,61,62,63,64,65,66,67,68,69,70}. Following the technology development three of the articles published this year focused on the integration of the cloud and the IoT, the applications, the requirements and the security issues that arise from it \cite{54,56,60}.  Security was also one aspect that was covered from a number of survey articles \cite{56,63,70}. Authors in \cite{63} covered several aspects of IoT security, e.g. general devices security, communication security, network security, and application while in \cite{70} mechanisms that reassure secure routing were investigated. In contrast to previous years, surveys that published during 2016 covered new areas, such as SDN and virtualization \cite{66}, economic and pricing theory in IoT \cite{68}, social internet of vehicles \cite{69} and data quality \cite{61}. Other topics covered from the survey articles were middleware \cite{55}, context aware computing \cite{59}, applications of IoT in the healthcare industry \cite{64}, cognitive radio technology \cite{67}, data models \cite{65}, mobile crowd sensing strategies \cite{58},the deployment of IoT in smart environments \cite{57} and the main proposed architectures for IoT \cite{62}.  Xie et al. \cite{40} surveys the security of the Web of Things (WoT) which is aimed to provide any electronic item (smart cards, sensors, etc.) with a URL. 

Among the aforementioned surveys, the security and privacy issues that are related to the IoT was thoroughly covered and analyzed \cite{3,4,9,13,14,20,24,29,32,35,39,40,41,45,47,49,50,56,63,70}. As it is shown in Tab. \ref{tab:Tab2c} data authentication and integrity was only covered partially from He et al. \cite{52} while the rest of the articles did not cover this major security aspect.  In this article we tend to survey authentication protocols for the IoT in four environments, including: (1) Machine to machine communications (M2M), (2) Internet of Vehicles (IoV), (3) Internet of Energy (IoE), and (4) Internet of Sensors (IoS).  Based on this thorough analysis open issues and future directions are identified that combine both innovative research along with the application, through appropriate adaptation, of existing solutions from other fields. We believe that this study will help researchers focus on the important aspects of authentication issues in the IoT  area and will guide them towards their future research. 

\begin{figure}[!]
 \centering
 \includegraphics[width=0.6\linewidth]{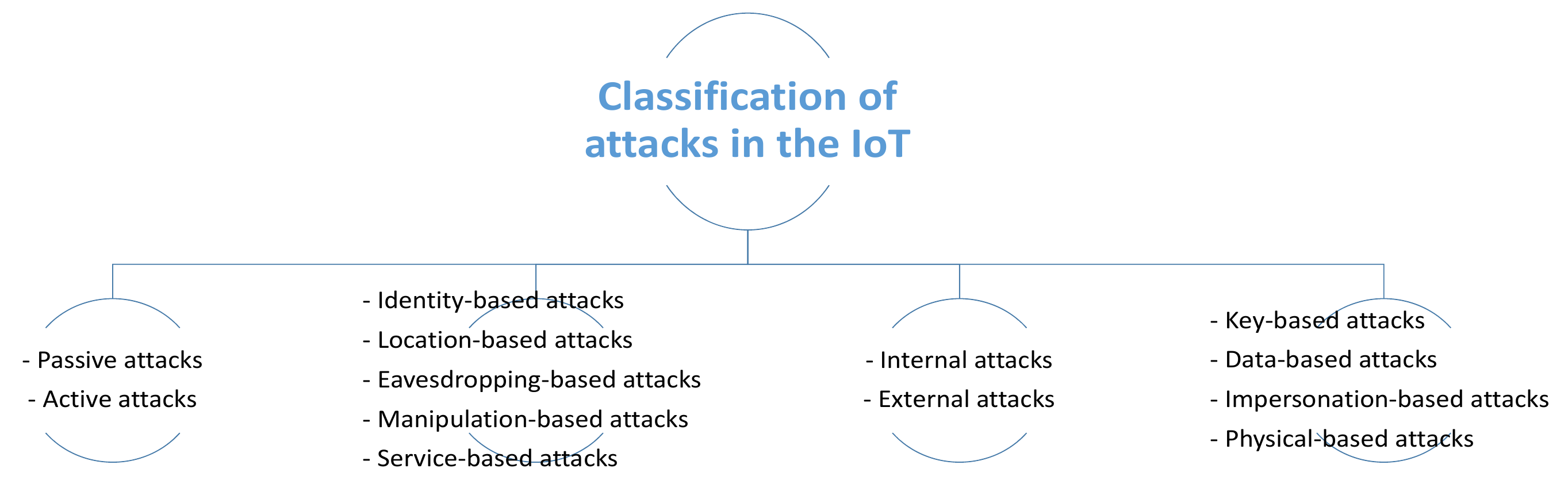}
 \caption{Classification of attacks in the IoT}
 \label{fig:Fig3a}
 \end{figure}
 
 \begin{table}[!]
 \centering
 \caption{Summary of attacks in Machine to machine communications (M2M) and defense protocols}
 {\tiny \checkmark :indicates fully supported; X: indicates not supported; 0: indicates partially supported. \\
 \begin{tabular}{|p{1.2in}|p{0.1in}|p{0.1in}|p{0.1in}|p{0.1in}|p{0.1in}|p{0.1in}|p{0.1in}|p{0.1in}|p{0.1in}|} \hline 
  & \multicolumn{9}{|p{2in}|}{\textbf{Authentication protocols for M2M}} \\ \hline 
 \textbf{Adversary model} & \cite{71} & \cite{76} & \cite{78} & \cite{79} & \cite{81} & \cite{117} & \cite{118} & \cite{124} & \cite{136} \\ \hline 
 Audio replay attack & \checkmark & 0 & X & 0 & 0 & 0 & X & X & 0 \\ \hline 
 Changing distance attack & \checkmark & X & X & X & X & X & X & X & X \\ \hline 
 Same-type-device attack & \checkmark & X & X & X & X & X & X & X & X \\ \hline 
 Composition attack & \checkmark & X & X & X & X & X & X & X & X \\ \hline 
 Redirection attack & 0 & \checkmark & 0 & \checkmark & X & X & 0 & X & \checkmark \\ \hline 
 Man-in-the-middle attack & 0 & \checkmark & 0 & \checkmark & 0 & 0 & X & X & \checkmark \\ \hline 
 Substitution attack & 0 & 0 & 0 & 0 & 0 & X & X & X & X \\ \hline 
 DoS attack & X & \checkmark & X & \checkmark & X & X & \checkmark & X & X \\ \hline 
 Replay attack & 0 & X & X & \checkmark & 0 & \checkmark & X & X & \checkmark \\ \hline 
 Forging attack & X & X & X & 0 & X & X & X & X & X \\ \hline 
 Colluding attack & 0 & X & X & 0 & X & X & 0 & X & X \\ \hline 
 Flooding attack & 0 & X & X & X & X & X & 0 & X & 0 \\ \hline 
 Side-channel attack & 0 & X & X & X & X & X & 0 & X & 0 \\ \hline 
 False messages attack & 0 & X & X & X & 0 & 0 & 0 & X & 0 \\ \hline 
 Sybil attack & X & X & X & X & 0 & 0 & X & X & 0 \\ \hline 
 Movement tracking & X & X & X & X & 0 & X & X & X & 0 \\ \hline 
 Message modification & X & X & X & X & 0 & X & X & X & X \\ \hline 
 Impersonation attack & X & X & X & X & 0 & \checkmark & \checkmark & X & X \\ \hline 
 Guessing attack & X & X & X & X & X & \checkmark & X & X & X \\ \hline 
 Stolen-verifier attack & X & X & X & X & X & \checkmark & X & X & X \\ \hline 
 Wormhole attack & 0 & 0 & X & 0 & X & 0 & X & X & 0 \\ \hline 
 Black hole attack & 0 & 0 & X & 0 & 0 & 0 & X & X & 0 \\ \hline 
 Attribute-trace attack & X & X & X & X & 0 & X & X & X & X \\ \hline 
 Eavesdropping attack & X & X & X & X & 0 & 0 & X & X & 0 \\ \hline 
 Chosen-plaintext attack & X & X & X & X & 0 & X & X & X & 0 \\ \hline 
 Spam attack & 0 & X & X & X & 0 & 0 & X & X & 0 \\ \hline 
 Identity theft attack & 0 & X & X & X & X & 0 & X & X & X \\ \hline 
 User manipulation attack & 0 & X & X & X & X & 0 & 0 & X & 0 \\ \hline 
 Routing attack & 0 & X & X & X & X & 0 & X & X & X \\ \hline 
 Linkability attack & 0 & X & X & X & X & X & X & X & X \\ \hline 
 Rejection attack & X & X & X & X & X & X & X & X & X \\ \hline 
 Successive-response attack & X & X & X & X & X & X & X & X & X \\ \hline 
 Packet analysis attack & X & 0 & X & X & X & 0 & X & X & 0 \\ \hline 
 Packet tracing attack & X & 0 & X & X & X & 0 & X & X & 0 \\ \hline 
 Brute-force attack & 0 & 0 & X & 0 & 0 & X & 0 & 0 & X \\ \hline 
 \end{tabular}}
 \label{tab:Tab3a}
 \end{table}
 \begin{table}[!]
 \centering
 \caption{Summary of attacks in Internet of Vehicles (IoV) and defense protocols}
 {\tiny \checkmark :indicates fully supported; X: indicates not supported; 0: indicates partially supported.\\
 \begin{tabular}{|p{1.2in}|p{0.1in}|p{0.1in}|p{0.1in}|p{0.1in}|p{0.1in}|p{0.1in}|p{0.1in}|p{0.1in}|p{0.1in}|} \hline 
  & \multicolumn{9}{|p{2in}|}{\textbf{Authentication protocols for IoV}} \\ \hline 
 \textbf{Adversary model} & \cite{89} & \cite{90} & \cite{91} & \cite{92} & \cite{93} & \cite{94} & \cite{95} & \cite{108} & \cite{125} \\ \hline 
 Audio replay attack & 0 & 0 & 0 & X & 0 & 0 & X & 0 & X \\ \hline 
 Changing distance attack & X & X & X & X & X & X & X & X & X \\ \hline 
 Same-type-device attack & X & X & X & X & X & X & X & X & X \\ \hline 
 Composition attack & X & X & X & X & X & X & X & X & X \\ \hline 
 Redirection attack & 0 & 0 & X & X & X & X & X & X & X \\ \hline 
 Man-in-the-middle attack & \checkmark & 0 & 0 & X & X & \checkmark & 0 & X & X \\ \hline 
 Substitution attack & 0 & 0 & 0 & X & X & 0 & \checkmark & X & X \\ \hline 
 DoS attack & \checkmark & X & X & \checkmark & \checkmark & \checkmark & X & X & X \\ \hline 
 Replay attack & \checkmark & \checkmark & \checkmark & X & 0 & 0 & 0 & \checkmark & 0 \\ \hline 
 Forging attack & 0 & \checkmark & X & X & X & 0 & X & X & X \\ \hline 
 Colluding attack & 0 & \checkmark & X & 0 & X & X & X & X & X \\ \hline 
 Flooding attack & X & X & X & 0 & X & X & X & X & X \\ \hline 
 Side-channel attack & X & X & X & 0 & \checkmark & X & X & X & X \\ \hline 
 False messages attack & X & X & X & X & \checkmark & X & X & X & 0 \\ \hline 
 Sybil attack & 0 & X & X & X & \checkmark & 0 & X & X & 0 \\ \hline 
 Movement tracking & X & X & X & X & X & X & X & \checkmark & X \\ \hline 
 Message modification & X & X & X & X & X & X & 0 & \checkmark & X \\ \hline 
 Impersonation attack & X & X & X & X & X & \checkmark & X & 0 & X \\ \hline 
 Guessing attack & X & X & X & X & X & X & X & X & 0 \\ \hline 
 Stolen-verifier attack & X & X & X & X & X & X & X & X & 0 \\ \hline 
 Wormhole attack & 0 & 0 & X & X & 0 & X & 0 & 0 & 0 \\ \hline 
 Black hole attack & 0 & 0 & X & X & 0 & X & 0 & 0 & 0 \\ \hline 
 Attribute-trace attack & X & X & X & X & X & 0 & X & X & 0 \\ \hline 
 Eavesdropping attack & X & X & 0 & 0 & 0 & X & X & 0 & 0 \\ \hline 
 Chosen-plaintext attack & X & X & X & 0 & X & X & 0 & X & 0 \\ \hline 
 Spam attack & X & X & X & 0 & X & 0 & 0 & X & X \\ \hline 
 Identity theft attack & X & X & X & 0 & X & X & 0 & X & X \\ \hline 
 User manipulation attack & X & X & X & 0 & X & X & 0 & 0 & X \\ \hline 
 Routing attack & 0 & X & 0 & X & 0 & X & 0 & 0 & 0 \\ \hline 
 Linkability attack & X & X & X & X & X & 0 & X & 0 & X \\ \hline 
 Rejection attack & X & X & X & X & X & 0 & X & 0 & 0 \\ \hline 
 Successive-response attack & X & X & X & X & X & 0 & X & X & X \\ \hline 
 Packet analysis attack & 0 & 0 & X & X & 0 & 0 & X & 0 & 0 \\ \hline 
 Packet tracing attack & 0 & 0 & X & X & 0 & 0 & X & 0 & 0 \\ \hline 
 Brute-force attack & X & X & X & X & X & 0 & X & 0 & 0 \\ \hline 
 \end{tabular}}
 \label{tab:Tab3b}
 \end{table}
 \begin{table}[!]
 \centering
 \caption{Summary of attacks in Internet of Energy (IoE) and defense protocols}
 {\tiny \checkmark :indicates fully supported; X: indicates not supported; 0: indicates partially supported.\\
 \begin{tabular}{|p{1.2in}|p{0.1in}|p{0.1in}|p{0.1in}|p{0.1in}|p{0.1in}|p{0.1in}|p{0.1in}|p{0.1in}|p{0.1in}|} \hline 
  & \multicolumn{9}{|p{2in}|}{\textbf{Authentication protocols for IoE}} \\ \hline 
 \textbf{Adversary model} & \cite{127} & \cite{128} & \cite{129} & \cite{130} & \cite{131} & \cite{132} & \cite{133} & \cite{134} & \cite{135} \\ \hline 
 Audio replay attack & X & X & X & X & X & X & X & X & X \\ \hline 
 Changing distance attack & 0 & X & X & X & X & 0 & 0 & 0 & X \\ \hline 
 Same-type-device attack & X & X & X & 0 & X & X & X & X & X \\ \hline 
 Composition attack & X & X & X & X & X & X & X & X & X \\ \hline 
 Redirection attack & X & X & X & 0 & X & 0 & X & X & X \\ \hline 
 Man-in-the-middle attack & 0 & 0 & 0 & \checkmark & 0 & 0 & \checkmark & 0 & 0 \\ \hline 
 Substitution attack & X & 0 & X & X & X & X & 0 & 0 & X \\ \hline 
 DoS attack & X & X & 0 & \checkmark & X & 0 & \checkmark & X & 0 \\ \hline 
 Replay attack & 0 & \checkmark & 0 & \checkmark & \checkmark & \checkmark & \checkmark & 0 & \checkmark \\ \hline 
 Forging attack & \checkmark & 0 & 0 & 0 & 0 & X & X & X & X \\ \hline 
 Colluding attack & X & 0 & X & 0 & 0 & X & 0 & 0 & X \\ \hline 
 Flooding attack & X & 0 & X & 0 & X & X & 0 & 0 & 0 \\ \hline 
 Side-channel attack & X & X & X & X & X & 0 & 0 & 0 & X \\ \hline 
 False messages attack & 0 & \checkmark & 0 & 0 & 0 & 0 & 0 & 0 & \checkmark \\ \hline 
 Sybil attack & 0 & 0 & 0 & 0 & 0 & 0 & X & X & 0 \\ \hline 
 Movement tracking & 0 & X & X & X & X & 0 & X & X & 0 \\ \hline 
 Message modification & 0 & \checkmark & 0 & 0 & 0 & 0 & 0 & 0 & \checkmark \\ \hline 
 Impersonation attack & 0 & 0 & X & X & 0 & X & 0 & 0 & 0 \\ \hline 
 Guessing attack & X & 0 & X & 0 & X & X & X & X & X \\ \hline 
 Stolen-verifier attack & X & X & X & X & X & X & X & X & X \\ \hline 
 Wormhole attack & X & X & 0 & X & X & 0 & 0 & 0 & 0 \\ \hline 
 Black hole attack & X & X & 0 & X & X & 0 & 0 & 0 & 0 \\ \hline 
 Attribute-trace attack & X & X & X & 0 & X & 0 & X & X & X \\ \hline 
 Eavesdropping attack & 0 & 0 & 0 & 0 & 0 & 0 & 0 & 0 & 0 \\ \hline 
 Chosen-plaintext attack & X & X & X & 0 & X & \checkmark & X & X & X \\ \hline 
 Spam attack & X & X & X & 0 & X & X & X & X & X \\ \hline 
 Identity theft attack & X & X & 0 & 0 & 0 & X & 0 & 0 & 0 \\ \hline 
 User manipulation attack & X & X & X & X & 0 & X & X & X & 0 \\ \hline 
 Routing attack & X & X & 0 & 0 & X & X & X & X & X \\ \hline 
 Linkability attack & 0 & X & 0 & 0 & X & X & 0 & 0 & X \\ \hline 
 Rejection attack & 0 & X & 0 & 0 & 0 & X & 0 & 0 & 0 \\ \hline 
 Successive-response attack & 0 & X & X & 0 & X & X & X & X & 0 \\ \hline 
 Packet analysis attack & 0 & \checkmark & 0 & 0 & 0 & X & 0 & 0 & \checkmark \\ \hline 
 Packet tracing attack & 0 & 0 & X & 0 & 0 & 0 & 0 & 0 & 0 \\ \hline 
 Brute-force attack & X & X & X & \checkmark & X & X & \checkmark & 0 & X \\ \hline 
 \end{tabular}}
 \label{tab:Tab3c}
 \end{table}
 \begin{table}[!]
 \centering
 \caption{Summary of attacks in Internet of Sensors (IoS) and defense protocols}
 {\tiny \checkmark :indicates fully supported; X: indicates not supported; 0: indicates partially supported.\\
 \begin{tabular}{|p{1.2in}|p{0.1in}|p{0.1in}|p{0.1in}|p{0.1in}|p{0.1in}|p{0.1in}|p{0.1in}|p{0.1in}|p{0.1in}|p{0.1in}|p{0.1in}|p{0.1in}|p{0.1in}|p{0.1in}|p{0.1in}|p{0.1in}|} \hline 
  & \multicolumn{16}{|p{2in}|}{\textbf{Authentication protocols for IoS}} \\ \hline 
 \textbf{Adversary model} & \cite{167} & \cite{168} & \cite{169} & \cite{170} & \cite{171} &  \cite{172} & \cite{173} & \cite{174} & \cite{175} & \cite{176} & \cite{177} & \cite{178} & \cite{179} & \cite{180} & \cite{181} & \cite{182} \\ \hline 
 Audio replay attack & X & X & X & X & X & X & X & X & X & X & X & X & X & X & X & X \\ \hline 
 Changing distance attack & 0 & X & 0 & X & X & X & X & X & X & X & X & X & X & X & X & X \\ \hline 
 Same-type-device attack & 0 & X & 0 & X & X & X & X & X & 0 & X & X & X & X & X & X & X \\ \hline 
 Composition attack & \checkmark & 0 & X & X & 0 & 0 & X & 0 & 0 & X & X & X & 0 & 0 & 0 & 0 \\ \hline 
 Redirection attack & \checkmark & 0 & 0 & 0 & X & 0 & 0 & 0 & 0 & 0 & 0 & 0 & 0 & 0 & 0 & 0 \\ \hline 
 Man-in-the-middle attack & 0 & 0 & 0 & 0 & 0 & \checkmark & 0 & 0 & \checkmark & 0 & 0 & 0 & \checkmark & \checkmark & \checkmark & 0 \\ \hline 
 Substitution attack & 0 & X & X & X & X & X & 0 & X & 0 & 0 & 0 & 0 & 0 & X & X & X \\ \hline 
 DoS attack & 0 & 0 & 0 & X & 0 & X & 0 & X & \checkmark & 0 & 0 & X & 0 & 0 & 0 & 0 \\ \hline 
 Replay attack & \checkmark & 0 & \checkmark & \checkmark & 0 & \checkmark & \checkmark & 0 & \checkmark & \checkmark & \checkmark & X & \checkmark & 0 & 0 & \checkmark \\ \hline 
 Forging attack & 0 & \checkmark & 0 & X & 0 & \checkmark & 0 & 0 & 0 & 0 & 0 & X & 0 & \checkmark & \checkmark & 0 \\ \hline 
 Colluding attack & 0 & 0 & 0 & X & 0 & 0 & 0 & X & 0 & 0 & 0 & \checkmark & 0 & 0 & 0 & 0 \\ \hline 
 Flooding attack & \checkmark & 0 & X & X & 0 & 0 & 0 & X & 0 & 0 & 0 & 0 & 0 & 0 & 0 & 0 \\ \hline 
 Side-channel attack & X & 0 & X & X & X & X & X & X & X & X & X & X & 0 & X & X & X \\ \hline 
 False messages attack & 0 & X & 0 & 0 & 0 & 0 & 0 & 0 & 0 & 0 & 0 & 0 & 0 & 0 & 0 & 0 \\ \hline 
 Sybil attack & 0 & 0 & \checkmark & 0 & X & X & X & 0 & 0 & 0 & 0 & 0 & 0 & 0 & 0 & 0 \\ \hline 
 Movement tracking & 0 & 0 & X & X & 0 & X & X & 0 & 0 & 0 & 0 & 0 & 0 & 0 & 0 & 0 \\ \hline 
 Message modification & 0 & 0 & 0 & 0 & 0 & 0 & 0 & \checkmark & 0 & 0 & 0 & 0 & \checkmark & 0 & 0 & 0 \\ \hline 
 Impersonation attack & \checkmark & \checkmark & 0 & \checkmark & \checkmark & 0 & 0 & \checkmark & 0 & \checkmark & \checkmark & X & \checkmark & 0 & 0 & \checkmark \\ \hline 
 Guessing attack & \checkmark & \checkmark & 0 & \checkmark & 0 & 0 & 0 & 0 & 0 & \checkmark & \checkmark & X & \checkmark & 0 & \checkmark & 0 \\ \hline 
 Stolen-verifier attack & \checkmark & X & X & 0 & 0 & X & X & X & \checkmark & 0 & 0 & 0 & \checkmark & 0 & 0 & 0 \\ \hline 
 Wormhole attack & 0 & 0 & 0 & 0 & 0 & 0 & 0 & X & X & X & X & X & 0 & X & X & X \\ \hline 
 Black hole attack & 0 & 0 & 0 & 0 & 0 & 0 & 0 & X & X & X & X & X & 0 & X & X & X \\ \hline 
 Attribute-trace attack & X & X & X & X & X & 0 & X & X & 0 & X & X & X & 0 & X & X & X \\ \hline 
 Eavesdropping attack & 0 & 0 & 0 & 0 & 0 & 0 & 0 & X & 0 & 0 & 0 & 0 & 0 & 0 & 0 & 0 \\ \hline 
 Chosen-plaintext attack & X & X & X & X & X & X & X & X & X & X & X & X & X & 0 & 0 & 0 \\ \hline 
 Spam attack & X & X & X & 0 & X & X & 0 & X & 0 & X & X & X & X & 0 & 0 & 0 \\ \hline 
 Identity theft attack & 0 & 0 & 0 & X & X & X & 0 & X & 0 & X & X & X & 0 & 0 & 0 & 0 \\ \hline 
 User manipulation attack & 0 & 0 & X & 0 & 0 & X & 0 & 0 & 0 & 0 & 0 & 0 & 0 & 0 & 0 & 0 \\ \hline 
 Routing attack & 0 & 0 & 0 & 0 & 0 & X & 0 & X & 0 & X & X & X & 0 & X & X & X \\ \hline 
 Linkability attack & 0 & 0 & 0 & 0 & 0 & 0 & 0 & X & 0 & X & X & X & 0 & X & X & X \\ \hline 
 Rejection attack & 0 & 0 & 0 & 0 & 0 & X & 0 & X & 0 & X & X & X & 0 & X & X & X \\ \hline 
 Successive-response attack & \checkmark & X & 0 & X & 0 & X & X & X & 0 & X & X & X & X & X & X & X \\ \hline 
 Packet analysis attack & 0 & 0 & X & 0 & 0 & X & X & \checkmark & X & X & X & X & X & 0 & 0 & 0 \\ \hline 
 Packet tracing attack & 0 & 0 & X & 0 & \checkmark & X & X & \checkmark & X & X & X & X & X & 0 & 0 & 0 \\ \hline 
 Brute-force attack & X & X & X & X & X & X & X & X & X & X & X & X & X & X & X & X \\ \hline 
 \end{tabular}}
 \label{tab:Tab3d}
 \end{table}
 
\section{Threat models}\label{sec:threat-models}
In this section, we discuss the threat models in the IoT. The summary of thirty-five attacks in M2M, IoV, IoE, IoS and defense protocols are given in Tab. \ref{tab:Tab3a}, Tab. \ref{tab:Tab3b}, Tab. \ref{tab:Tab3c}, and Tab. \ref{tab:Tab3d}, respectively. We focus on five attacks, which are mostly used by authors that propose new authentications protocols for evaluating their methods, namely, man-in-the-middle attack, impersonation attack, forging attack, and replay attack. Generally, the classification of attacks \cite{271,272,273,274} frequently mentioned in the literature is done using the following four types, as shown in Fig. \ref{fig:Fig3a}: 

\begin{enumerate}
\item  Type A: Passive or active;
\item  Type B: Internal or external;
\item  Type C \cite{149}: Key-based attacks, data-based attacks, impersonation-based attacks, and physical-based attacks;
\item  Type D \cite{233}: Identity-based attacks, location-based attacks, eavesdropping-based attacks, manipulation-based attack, and service-based attacks.
\end{enumerate}

\begin{figure}[!]
  \centering
  \includegraphics[width=0.4\linewidth]{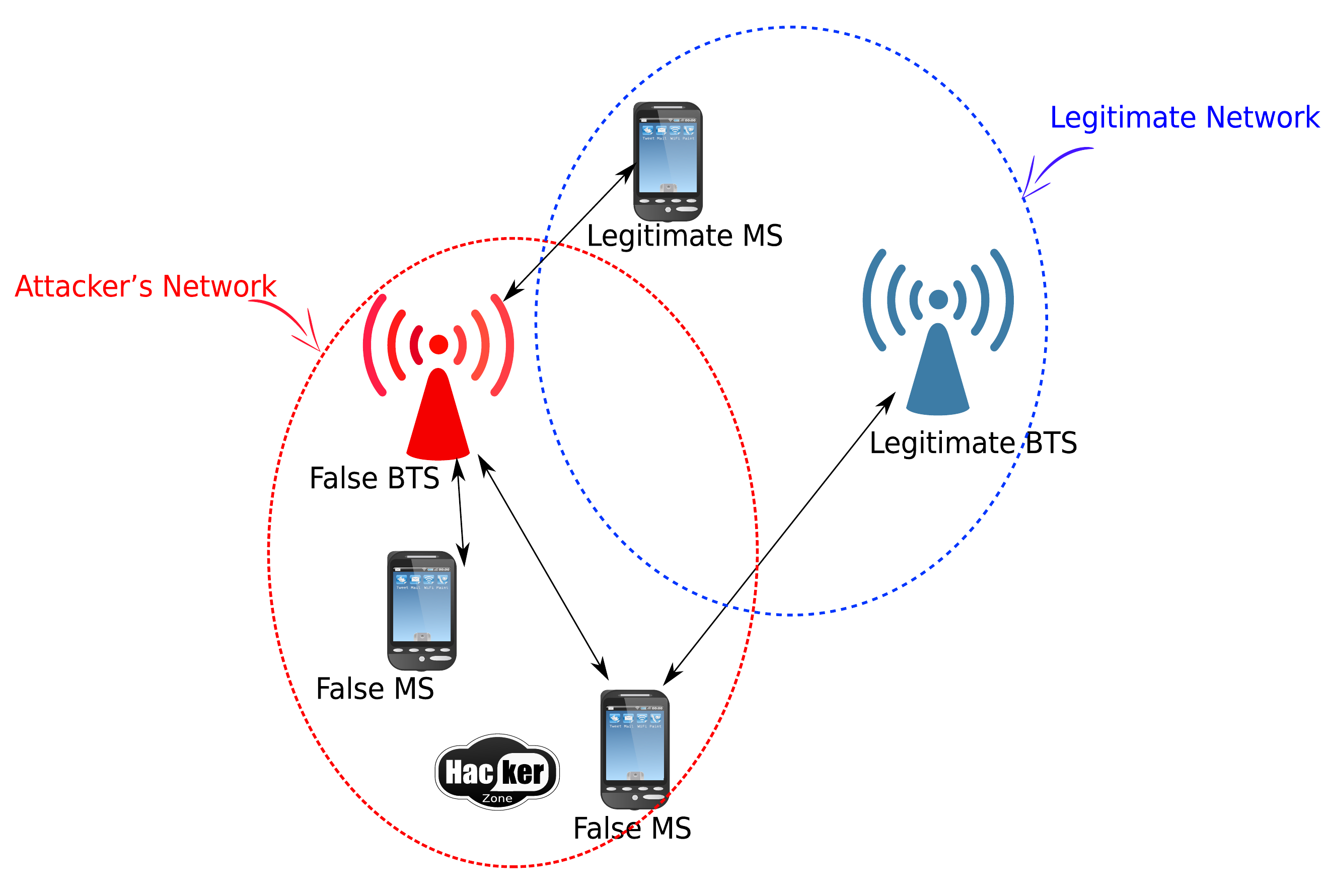}
  \caption{MITM attack on GSM as defined by Conti et al. in \cite{275}, BTS : Base Transceiver Station; MS: Mobile Station}
  \label{fig:Fig3b}
  \end{figure}
  
\subsection{Man-in-the-middle attack}

The Man-In-The-Middle (MITM) attack is one of the most well known attacks in the IoT. With MITM attack, an adversary can spoof the identities of two honest nodes (N1 and N2) involved in a network exchange and pass N1 for N2 and vice versa, i.e., takes control of the communication channel between N1 and N2. Under this control, an adversary can intercept, modify, change, or replace target victims' communication traffic. However, we note here that there is a good survey article published in 2016 by Conti et al. in \cite{275}, which presents a comprehensive survey on MITM attacks. Specifically, authors in \cite{275} classify MITM attacks in three different categories, namely, 1) MITM based on impersonation techniques, 2) MITM based on the communication channel, and 3) MITM based on the location of an adversary. As presented in Fig. \ref{fig:Fig3b}, at any moment an adversary can set-up a connection between False BTS and Legitimate MS, where False MS impersonates the victim's MS to the real network by resending the identity information. Moreover, as presented in Tab. \ref{tab:Tab3e}, there are twelve authentication protocols for the IoT, which can detect and avoid the MITM attack.  The four authentication protocols in \cite{76,175,179,180} use the idea of mutual authentication. The two authentication protocols \cite{79} and \cite{136} use the idea of authentication acknowledge phase. With the protocol \cite{130}, all packets are fully encrypted with the receiver's public key, which can prevent the MITM attack. On the other hand, with the protocol \cite{89}, when the keys generated at the mobile router and the relay router for authentication are based on the concept of symmetric polynomials, an adversary can not identify a shared key between two legitimate users making it impossible for him to impersonate a mobile router or a relay router. In addition, both protocols \cite{133} and \cite{172} are based on a password and biometric update phase in order to prevent that an adversary can impersonate the passwords of a smart meter.

\subsection{Impersonation and forging attack}

Under the impersonation and forging attack in the IoS, an adversary can eavesdrop or intercept the login request message of  previous sessions over the public/open channel during authentication protocol execution. After that, he can modify and re-transmit the message to the user in order to impersonate as a valid user, as defined by Amin et al. \cite{170}. We note that this attack is analyzed more in authentication protocols that are produced for the IoS. Moreover, as presented in Tab. \ref{tab:Tab3f} there are sixteen authentication protocols for the IoT, which can detect the impersonation and forging attack. The protocol \cite{90} uses two ideas, namely, 1) linear search algorithm and 2) binary search algorithm. The protocol \cite{118} uses strong anonymous access authentication and user tracking on a disputed access request, to prevent the impersonation and forging attack. Besides, the idea of using a password for detecting the impersonation of the gateway node is presented by four authentication protocols \cite{179,182,117}, and \cite{181}. In addition, the hash mechanism which is applied on the shared key between gateway wireless node and sensors can prevent the impersonation of a sensor.
 
 \begin{table}[!]
 \centering
 \caption{Approaches for detecting and avoiding the man-in-the-middle attack}
 {\tiny
 \begin{tabular}{|p{1.4in}|p{3in}|p{2.5in}|} \hline 
 \textbf{Protocol} & \textbf{Data attacked} & \textbf{Approach} \\ \hline 
 Lai et al. (2016) \cite{76} & - Communication channel between the mobile management entity and the home subscriber server & - Mutual authentication and key agreement between multiple M2M devices and the core network simultaneously \\ \hline 
 Lai et al. (2013) \cite{79} & - The data between the mobiles equipment's and the 3GPP network & - Authentication acknowledge phase \\ \hline 
 Céspedes et al. (2013) \cite{89} & - Identify a shared key between two legitimate users;\newline - Impersonate a mobile router or a relay router; & - The keys generated at the mobile router and the relay router for authentication are based on the concept of symmetric polynomials \\ \hline 
 Dolev et al. (2016) \cite{94} & - Communication channel between the vehicles & - Twofold authentication;\newline - Periodic certificate restore; \\ \hline 
 Nicanfar et al. (2011) \cite{130} & - Communication channel between the smart meter and the authentication agent;\newline - Communication channel between the authentication agent and the security associate (SA) server & - All packets are fully encrypted with the receivers public key \\ \hline 
 Nicanfar et al. (2014) \cite{133} & - The passwords of smart meter & - Changing the server password more often \\ \hline 
 Das (2016) \cite{172} & - The login request message during the login phase & - Password and biometric update phase \\ \hline 
 Lai et al. (2013) \cite{136} & - Can occur while connecting to a base station & - Authentication acknowledge phase \\ \hline 
 Farash et al. (2016) \cite{175} & - Data between the sensor node, users and gateway node & - Mutual authentication \\ \hline 
 Jiang et al. (2016) \cite{179} & - Data between the Sensor node, users and Gateway node & - Mutual authentication \\ \hline 
 Wu et al. (2016) \cite{180} & - Data between the Sensor node, users and Gateway node & - Mutual authentication \\ \hline 
 Das et al. (2016) \cite{181} & - The lost/stolen smart card of a legal user & - Password change phase \\ \hline 
 \end{tabular}}
 \label{tab:Tab3e}
 \end{table}
 \begin{figure}[h]
    \centering
    \includegraphics[width=0.25\linewidth]{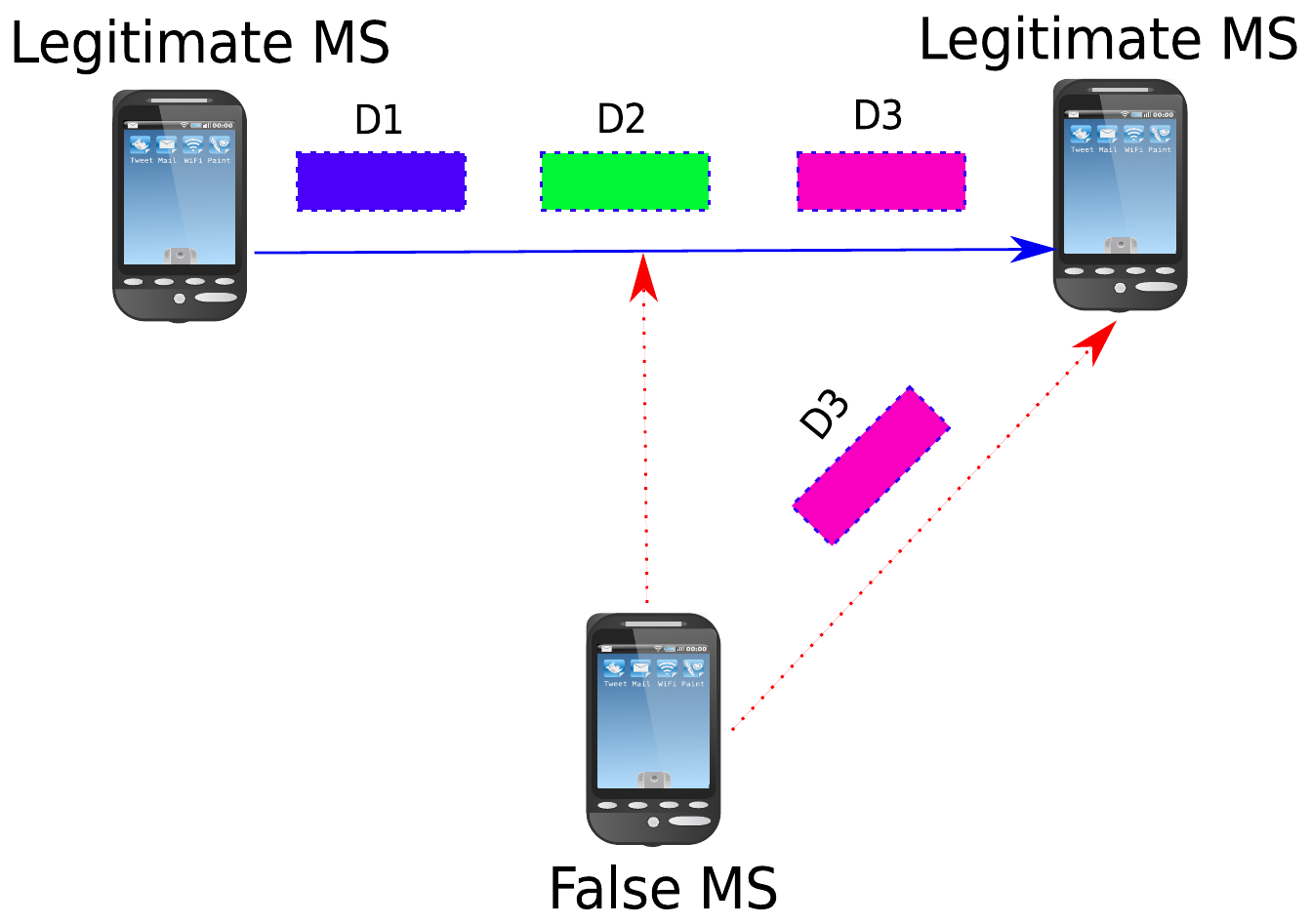}
    \caption{Replay attack, MS: Mobile Station}
    \label{fig:Fig3c}
    \end{figure}
 \begin{table}[!]
 \centering
 \caption{Approaches for detecting and avoiding the impersonation and forging attack}
 {\tiny
 \begin{tabular}{|p{1.4in}|p{2.5in}|p{3in}|} \hline 
 \textbf{Protocol} & \textbf{Data attacked} & \textbf{Approach} \\ \hline 
 Wasef et al. (2013) \cite{90} & - Forge the revocation check & - Linear search algorithm\newline - Binary search algorithm \\ \hline 
 Chung et al. (2016) \cite{168} & - Impersonate the mobile node & - Login and authentication phase \\ \hline 
 Das (2016) \cite{172}\newline  & - Eavesdrop or intercept the login request message of the previous sessions & - Authentication and key agreement phase \\ \hline 
 Wu et al. (2016) \cite{180} & - The data produced by the smart card in the Login phase & - Elliptic curve cryptosystem \\ \hline 
 Das et al. (2016) \cite{181} & - Eavesdrop, modify, or delete the contents of the transmitted messages  & - Password and biometric update \\ \hline 
 Sun et al. (2015) \cite{117} & - Information leakage of the M2M server & - The authentication process based on password \\ \hline 
 Lai et al. (2014) \cite{118} & - Forge and/or modifies the authentication messages & - Strong anonymous access authentication;\newline - User tracking on a disputed access request; \\ \hline 
 Dolev et al. (2016) \cite{94} & - Forge and/or modifies the authentication messages & - Two rounds of session key \\ \hline 
 Kumari et al. (2016) \cite{167} & - Impersonation of user and sensor node & - Gateway wireless node does not maintain any record to store user-specific information \\ \hline 
 Amin et al. (2016) \cite{170} & - Intercepts the login request message  & - Authentication and key agreement \\ \hline 
 Gope et al. (2016) \cite{171} & - The server's secret key & - Adversary has no knowledge about the secret identity of the gateway \\ \hline 
 Jiang et al. (2016) \cite{174} & - Gets the user smart card & - The hash mechanism using the shared key between gateway wireless node and sensor \\ \hline 
 Srinivas et al. (2017) \cite{176} & - Impersonation of the gateway node & - Non-invertible cryptographic one way hash function property \\ \hline 
 Kumari et al. (2016) \cite{177} & - Impersonation of the gateway node & - Secret session key \\ \hline 
 Jiang et al. (2016) \cite{179} & - Gets the user smart card & - Password \\ \hline 
 Liu and Chung (2016) \cite{182} & - Intercepts the login request message & - Password \\ \hline 
 \end{tabular}}
 \label{tab:Tab3f}
 \end{table}
 \begin{table}[!]
 \centering
 \caption{Approaches for detecting and avoiding the replay attack}
 {\tiny
 \begin{tabular}{|p{1.5in}|p{3in}|p{1.2in}|} \hline 
 \textbf{Protocol} & \textbf{Data attacked} & \textbf{Approach} \\ \hline 
 Lai et al. (2013) \cite{79} & - Replaying the data between the mobiles equipment's and the 3GPP network & - Random numbers \\ \hline 
 Sun et al. (2015) \cite{117} & - Replaying the intercepted login message & - Random numbers \\ \hline 
 Lai et al. (2013) \cite{136} & - Replaying the message between serving gateway and home subscriber server. & - Random numbers \\ \hline 
 Céspedes et al. (2013) \cite{89} & - Replaying one of the router solicitation messages & - Random numbers \\ \hline 
 Wasef et al. (2013) \cite{90} & - Replaying the disseminated messages in IoV & - Timestamp \\ \hline 
 Shao et al. (2016) \cite{91} & - Replaying the disseminated messages in IoV & - Timestamp \\ \hline 
 Zhang et al. (2015) \cite{108} & - Replaying the disseminated messages in IoV & - Timestamp \\ \hline 
 Li et al. (2014) \cite{128} & - Replaying the electricity consumption reports & - Merkle hash tree technique \\ \hline 
 Nicanfar et al. (2011) \cite{130} & - Replaying the electricity consumption reports & - Timestamp \\ \hline 
 Chim et al. (2011) \cite{131} & - Replaying the electricity consumption reports & - Timestamp \\ \hline 
 Fouda et al. (2011) \cite{132} & - Replaying the electricity consumption reports & - Timestamp \\ \hline 
 Nicanfar et al. (2014) \cite{133} & - Forwarding a previous acknowledgment from the smart meter to the server & - Timestamp \\ \hline 
 Mahmood et al. (2016) \cite{135} & - Intercept messages by home area network and replay those archaic messages to building area network gateway & - Timestamp \\ \hline 
 Kumari et al. (2016) \cite{167} & - Intercept and replay the login request to gateway wireless node & - Timestamp \\ \hline 
 Jan et al. (2016) \cite{169} & - Eavesdrop on advertisement packets and/or join-request packets and replay in other parts of the network. & - Hash function and ring keys \\ \hline 
 Amin et al. (2016) \cite{170} & - Replaying the message in the IoS & - Timestamp \\ \hline 
 Das (2016) \cite{172} & - Replaying the login request message\newline  & - Timestamp \\ \hline 
 Chang et al. (2016) \cite{173} & - Replaying the login request message\newline  & - Timestamp \\ \hline 
 Farash et al. (2016) \cite{175} & - Replaying the login request message\newline  & - Timestamp \\ \hline 
 Srinivas et al. (2017) \cite{176} & - Replaying the messages in the IoS\newline  & - Timestamp \\ \hline 
 Kumari et al. (2016) \cite{177} & - Intercept and replay the login request to gateway wireless node & - Timestamp \\ \hline 
 Jiang et al. (2016) \cite{179}\newline  & - Intercept the login request & - Timestamp \\ \hline 
 Liu and Chung \cite{182} & - Intercept the login request & - Timestamp \\ \hline
 \end{tabular}}
 \label{tab:Tab3g}
 \end{table}
 
\subsection{Replay attack}

The Replay attacks are MITM attacks, which consist of intercepting data packets and retransmitting them as is (without any decryption) to the destination server, as shown in Fig. \ref{fig:Fig3c} (intercepting D3 and retransmitting it). Under this attack, an adversary can obtain the same rights as the user. However, there are twenty-four authentication protocols for the IoT, which can detect and avoid the Replay attack, as presented in Tab. \ref{tab:Tab3g}. These authentication protocols use three ideas, namely, Timestamp, Hash function, and Random numbers. The idea of random numbers is used by \cite{79,117,136}, and \cite{89}. The idea of hash function is used by protocols \cite{128} and \cite{169}, such as the IPSec protocol implements an anti-replay mechanism based on Message Authentication Code (MAC) \cite{282}. In addition, the idea of Timestamp in the encrypted messages is used by \cite{90,91,108,128,130,131,132,133,135,167,169,170,172,173,175,176,177,179,182}.

\begin{figure}[!]
      \centering
      \includegraphics[width=0.5\linewidth]{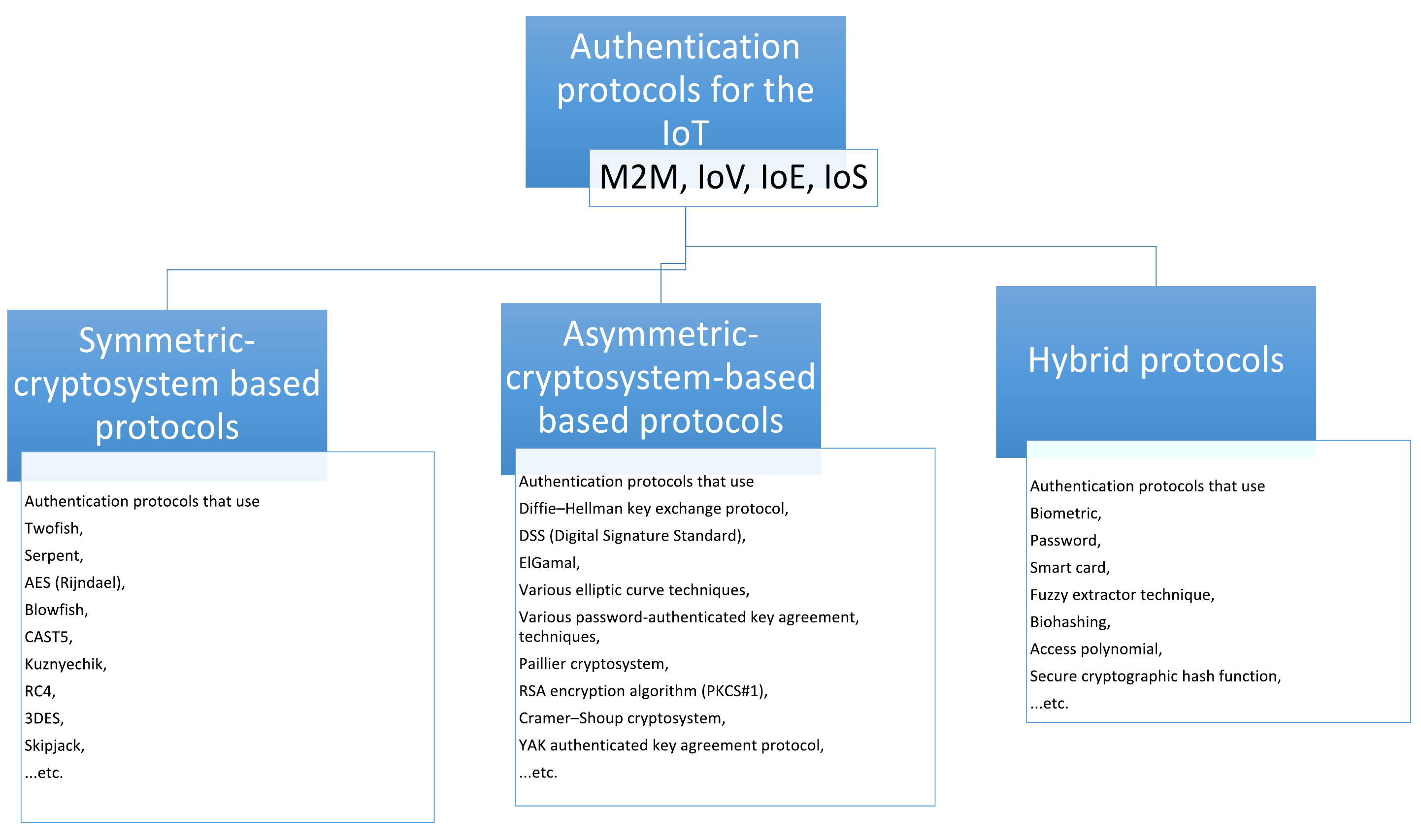}
      \caption{Classification of the existing authentication protocols for the IoT Based on the cryptosystems}
      \label{fig:Fig4a}
      \end{figure}
      
\section{Countermeasures and formal security verification techniques}\label{sec:countermeasures-and-formal-security-verification-techniques}
In order to satisfy the authentication model for secure IoT, namely, mutual authentication, perfect forward secrecy, anonymity, and untraceability, the authentication protocols use both cryptosystems and non-cryptosystems countermeasures. Tables \ref{tab:Tab4a}, \ref{tab:Tab4b}, \ref{tab:Tab4c}, and \ref{tab:Tab4d} present the cryptosystems and countermeasures used in authentication protocols for M2M, IoV, IoE, and IoS, respectively. In this section, we will discuss the countermeasures and present the formal security verification techniques used in these authentication protocols for the IoT.

\begin{table}[!]
\centering
\caption{Cryptosystems and Countermeasures used in Authentication protocols for Machine to machine communications (M2M)}
{\tiny
\begin{tabular}{|p{2.5in}|p{0.1in}|p{0.1in}|p{0.1in}|p{0.1in}|p{0.1in}|p{0.1in}|p{0.1in}|p{0.1in}|p{0.1in}|} \hline 
 & \multicolumn{9}{|p{2in}|}{\textbf{Authentication protocols for M2M}} \\ \hline 
\textbf{Cryptosystems \&  Countermeasures} & \cite{71} & \cite{76} & \cite{78} & \cite{79} & \cite{81} & \cite{117} & \cite{118} & \cite{124} & \cite{136} \\ \hline 
Secure cryptographic hash function \cite{155} &  & \checkmark & \checkmark & \checkmark & \checkmark & \checkmark &  & \checkmark & \checkmark \\ \hline 
Original data acquisition &  &  &  &  &  &  &  & \checkmark &  \\ \hline 
Spatial-Domain transformation &  &  &  &  &  &  &  & \checkmark &  \\ \hline 
Time-domain transformation &  &  &  &  &  &  &  & \checkmark &  \\ \hline 
Correlation coefficient-based matching algorithm (C-MA) & \checkmark &  &  &  &  &  &  &  &  \\ \hline 
Deviation ratio-based matching algorithm (D-MA) & \checkmark &  &  &  &  &  &  &  &  \\ \hline 
Aggregate message authentication codes (AMACs) \cite{80} &  & \checkmark &  &  &  &  &  &  & \checkmark \\ \hline 
Certificateless aggregate signature \cite{82} &  &  & \checkmark &  &  &  &  &  &  \\ \hline 
Elliptic Curve Diffie-Hellman (ECDH) \cite{85} &  &  &  & \checkmark &  &  &  &  &  \\ \hline 
ID-based signature scheme \cite{86} &  &  &  &  & \checkmark &  &  &  &  \\ \hline 
Advanced encryption standard (AES) \cite{120} &  &  &  &  &  & \checkmark &  &  &  \\ \hline 
Hybrid Linear Combination Encryption \cite{123} &  &  &  &  &  &  & \checkmark &  &  \\ \hline 
\end{tabular}}
\label{tab:Tab4a}
\end{table}

\begin{table}[!]
\centering
\caption{Cryptosystems and Countermeasures used in Authentication protocols for Internet of Vehicles (IoV)}
{\tiny
\begin{tabular}{|p{2.5in}|p{0.1in}|p{0.1in}|p{0.1in}|p{0.1in}|p{0.1in}|p{0.1in}|p{0.1in}|p{0.1in}|p{0.1in}|} \hline 
 & \multicolumn{9}{|p{2in}|}{\textbf{Authentication protocols for IoV}} \\ \hline 
\textbf{Cryptosystems \&  Countermeasures} & \cite{89} & \cite{90} & \cite{91} & \cite{92} & \cite{93} & \cite{94} & \cite{95} & \cite{108} & \cite{125} \\ \hline 
Secure cryptographic hash function \cite{155} &  & \checkmark & \checkmark & \checkmark & \checkmark & \checkmark &  & \checkmark & \checkmark \\ \hline 
Proxy Mobile IP (PMIP) \cite{97} & \checkmark &  &  &  &  &  &  &  &  \\ \hline 
Symmetric Polynomials \cite{98} & \checkmark &  &  &  &  &  &  &  &  \\ \hline 
Search Algorithms \cite{99} &  & \checkmark &  &  &  &  &  &  &  \\ \hline 
Group Signature \cite{100} \cite{101} &  &  & \checkmark &  &  &  &  &  &  \\ \hline 
Merkle Hash Tree (MHT) \cite{106} &  &  &  & \checkmark &  &  &  &  &  \\ \hline 
TESLA Scheme \cite{103,104,105} &  &  &  & \checkmark &  &  &  &  &  \\ \hline 
ECDSA Signature \cite{107} &  &  &  & \checkmark &  &  &  &  &  \\ \hline 
Multiplicative secret sharing technique \cite{109} &  &  &  &  & \checkmark &  &  &  &  \\ \hline 
Identity-based Public Key Cryptosystem \cite{111} &  &  &  &  &  &  &  & \checkmark &  \\ \hline 
Identity-based aggregate signature \cite{110} &  &  &  &  &  &  &  & \checkmark &  \\ \hline 
Digital signatures \cite{114} &  &  &  &  &  & \checkmark &  &  &  \\ \hline 
Anonymous attribute-based group setup scheme \cite{126} &  &  &  &  &  &  &  &  & \checkmark \\ \hline 
Keyed-hashing for message authentication (HMAC) \cite{243} &  &  &  &  &  &  & \checkmark &  &  \\ \hline 
\end{tabular}}
\label{tab:Tab4b}
\end{table}

\begin{table}[!]
\centering
\caption{Cryptosystems and Countermeasures used in Authentication protocols for Internet of Energy (IoE)}
{\tiny
\begin{tabular}{|p{2.5in}|p{0.1in}|p{0.1in}|p{0.1in}|p{0.1in}|p{0.1in}|p{0.1in}|p{0.1in}|p{0.1in}|p{0.1in}|} \hline 
 & \multicolumn{9}{|p{2in}|}{\textbf{Authentication protocols for IoE}} \\ \hline 
\textbf{Cryptosystems \&  Countermeasures} & \cite{127} & \cite{128} & \cite{129} & \cite{130} & \cite{131} & \cite{132} & \cite{133} & \cite{134} & \cite{135} \\ \hline 
 Secure cryptographic hash function \cite{155} & \checkmark &  & \checkmark &  &  & \checkmark & \checkmark &  &  \\ \hline 
HORS scheme \cite{148} & \checkmark &  &  &  &  &  &  &  &  \\ \hline 
 Heavy signing light verification (HSLV) \cite{148} & \checkmark &  &  &  &  &  &  &  &  \\ \hline 
 Light signing heavy verification (LSHV) \cite{148} & \checkmark &  &  &  &  &  &  &  &  \\ \hline 
 Merkle Hash tree technique \cite{150}  &  & \checkmark &  &  &  &  &  &  &  \\ \hline 
 Short signatures (BLS) \cite{152} &  &  & \checkmark &  &  &  &  &  &  \\ \hline 
 Batch verification \cite{153} &  &  & \checkmark &  &  &  &  &  &  \\ \hline 
 Signature aggregation \cite{154} &  &  & \checkmark &  &  &  &  &  &  \\ \hline 
Identity-based Public Key Cryptosystem \cite{111} &  &  &  & \checkmark &  &  &  &  &  \\ \hline 
 Public-key encryption, such as RSA \cite{156} &  &  &  &  & \checkmark &  &  & \checkmark &  \\ \hline 
 HMAC, such as SHA-1 \cite{157} and MD5 \cite{158} &  &  &  &  & \checkmark &  &  & \checkmark & \checkmark \\ \hline 
 Diffie-Hellman key establishment protocol \cite{159} &  &  &  &  &  & \checkmark &  &  &  \\ \hline 
 EIBC mechanism \cite{161} &  &  &  &  &  &  & \checkmark &  &  \\ \hline 
 ID-based cryptography (IBC) \cite{162} &  &  &  &  &  &  & \checkmark &  &  \\ \hline 
Digital signatures \cite{114} &  &  &  &  &  &  &  & \checkmark &  \\ \hline 
Homomorphic Encryption \cite{164} &  &  &  &  &  &  &  & \checkmark &  \\ \hline 
Bloom Filter \cite{165} &  &  &  &  &  &  &  & \checkmark &  \\ \hline 
Commitment scheme &  &  &  &  &  &  &  & \checkmark &  \\ \hline 
Symmetric encryption/decryption algorithm \cite{159} &  &  &  &  &  &  &  &  & \checkmark \\ \hline 
\end{tabular}}
\label{tab:Tab4c}
\end{table}

\subsection{Countermeasures}

Based on the cryptosystems, the existing authentication protocols for the IoT can mainly be classified into three categories: symmetric-cryptosystem based, asymmetric-cryptosystem-based, and hybrid protocols, as shown in Fig. \ref{fig:Fig4a}. As presented in the following tables (\ref{tab:Tab4a}, \ref{tab:Tab4b}, \ref{tab:Tab4c}, and \ref{tab:Tab4d}), most authentication protocols use a secure cryptographic hash function \cite{155}. 

As presented in Tab. \ref{tab:Tab4a}, the protocol \cite{124} uses three cryptosystems, namely, original data acquisition, spatial-domain transformation, and time-domain transformation. The protocol \cite{71} use two matching algorithms, namely, correlation coefficient-based matching algorithm (C-MA) and deviation ratio-based matching algorithm (D-MA). The aggregate message authentication codes (AMACs) \cite{80} is used by both schemes \cite{76} and \cite{136}. The AMAC tool is a tuple of the following probabilistic polynomial-time algorithms: \textit{Authentication algorithm}, \textit{Aggregation algorithm}, and \textit{Verification algorithm}. The Authentication algorithm outputs a $tag$ tag, where the aggregate of tags can be simply computing the XOR of all the tag values; i.e., $tag={tag}_1\bigoplus {tag}_2\bigoplus \cdots \bigoplus {tag}_l$, where $1,\ .\ .\ .\ ,l$ are identifiers. The protocol \cite{78} uses certificateless aggregate signature \cite{82}, which enables an algorithm to aggregate $n$ signatures of $n$ distinct messages from $n$ users into a single short signature. In addition, the certificateless aggregate signature scheme is secure against existential forgery in the chosen aggregate model.  For an aggregating set of $n$ users $\{U_1,\dots ,U_n\}$ with identities $\{{ID}_1,\dots ,{ID}_n\}$ and the corresponding public keys $\{{upk}_1,\dots ,{upk}_n\}$, and message-signature pairs $\left(m_1,{\sigma }_1=\left(U_1,V_1\right)\right),\dots ,\left(m_n,{\sigma }_n=\left(U_n,V_n\right)\right)\ $ from $\{U_1,\dots ,U_n\}$ respectively, the aggregate signature generator computes $V=\sum^n_{i=1}{V_i}$ and outputs ${\sigma }_n=(U_1,\dots ,U_n,\ V)$ as an aggregate signature. The protocol \cite{79} use Elliptic Curve Diffie-Hellman (ECDH) \cite{85}, which is an anonymous key agreement protocol. The protocol \cite{81} uses ID-based signature scheme \cite{86} that consists of four algorithms, \textit{Setup}, \textit{Extract}, \textit{Sign}, and \textit{Verify}. With \textit{Setup} algorithm, the trust authority chooses efficiently computable monomorphisms. The trust authority performs the \textit{Extract} algorithm when a signer requests the secret key corresponding to their identity. The $Sign$ algorithm produce a signature from the user with identity $ID$ on the message $m$. Therefore, the protocol \cite{117} uses advanced encryption standard (AES) \cite{120}, which is a symmetric encryption standard intended to replace the Data Encryption Standard (DES) \cite{240} that has become too weak in view of current attacks. The protocol \cite{118} uses the Linear Combination Encryption (LCE) \cite{123}, which is an extension of ElGamal encryption \cite{241} that is secure in groups where the Decision Diffie--Hellman (DDH) problem is easy but the Computational Diffie--Hellman (CDH) problem is hard. With the LCE scheme \cite{123}, a user's public and secret keys are defined as $pk=\left(u,v,w_1=u^x,{\ w}_2=v^y\right)$ and $sk=(x,y)$, where $u,v\leftarrow G_1$ and $x,y\leftarrow Z^*_p$. The message $M$ is encrypted to $(D_1=u^a,\ D_2=v^b,\ D_3=M.w^a_1w^b_2)$ where $a,b\in Z^*_p$ are random. Then, the original message $M$ is decrypted from the ciphertext $(D_1,D_2,D_3)$ by $D_3.{(D^x_1.D^y_2)}^{-1}$. 

As presented in Tab. \ref{tab:Tab4b}, the protocol \cite{89} uses both countermeasures, namely, Proxy Mobile IP (PMIP) \cite{97} and Symmetric Polynomials \cite{98}. The PMIP is a localized network-based IP mobility protocol (RFC 5213 \cite{242}) that defines two entities: the Mobile Access Gateway (MAG) and the Local Mobility Anchor (LMA). The symmetric polynomial is defined as any polynomial of two or more variables that achieves the interchangeability property, i.e., $f(x,y)=f(y,x)$. For example, given two users identities 1 and 2, and the symmetric polynomial $f\left(x,y\right)=x^2y^2+xy+10$, the resultant evaluation functions are $f\left(1,y\right)=y^2+y+10$ and $f\left(2,y\right)=4y^2+2y+10$, respectively. Then, if user 1 evaluates its function $\ f\left(1,y\right)$ for user 2, it obtains $f\left(1,2\right)=16$. In the same way, $f\left(2,y\right)$ for user 1, user 2 obtains $f\left(1,2\right)=16$. As result, both users share a secret key, 16, without transmitting any additional messages to each other. Contrary to this idea of symmetric polynomials, the protocol \cite{90} uses the idea of search algorithms \cite{99}, which include non-optimized search algorithms such as linear search algorithm, and optimized search algorithms such as binary search algorithm, and lookup hash tables. In another work \cite{100} Chaum and van Heyst  introduce the idea of group signatures in order to providing anonymity for signers. The protocol \cite{91} uses this idea based on the Strong Diffie-Hellman assumption and the Decision Linear assumption. The protocol \cite{92} uses three countermeasures, namely, 1) Merkle Hash Tree (MHT) \cite{106}, 2) TESLA scheme \cite{103,104,105}, and 3) Elliptic Curve Digital Signature Algorithm (ECDSA) \cite{107}. The MHT is a binary tree structure where each leaf is assigned a hash value and an inner node is assigned the hash value of its children. To achieve source authentication, the TESLA scheme uses one-way hash chains with the delayed disclosure of keys based on symmetric cryptography. The protocol \cite{93} uses multiplicative secret sharing technique \cite{109} where the user can generate one-time pseudonym private key pairs and leakage-resilient locally. Similar to the protocol \cite{91}, the protocol \cite{94} uses the idea of digital signatures \cite{114}. The protocol \cite{95} uses keyed-hashing for message authentication (HMAC) \cite{243} to instantiate the pseudorandom function in the prototype implementation of electric vehicle ecosystem. The protocol \cite{108} uses two similar ideas, namely, identity-based public key cryptosystem \cite{111} and identity-based aggregate signature \cite{110}. For providing a flexible attribute management, the protocol \cite{125} uses an anonymous attribute-based group setup scheme \cite{126} that incorporates the policy-based data access control in the ciphertext.

\begin{table}[!]
\centering
\caption{Cryptosystems and Countermeasures used in Authentication protocols for Internet of Sensors (IoS)}
{\tiny
\begin{tabular}{|p{2in}|p{0.1in}|p{0.1in}|p{0.1in}|p{0.1in}|p{0.1in}|p{0.1in}|p{0.1in}|p{0.1in}|p{0.1in}|p{0.1in}|p{0.1in}|p{0.1in}|p{0.1in}|p{0.1in}|} \hline 
 & \multicolumn{14}{|p{3.7in}|}{\textbf{Authentication protocols for IoS}} \\ \hline 
\textbf{Cryptosystems \&  Countermeasures} & \cite{167} & \cite{168} & \cite{169} & \cite{170} & \cite{171} & \cite{172} & \cite{173} & \cite{174} & \cite{175} & \cite{176} & \cite{177} & \cite{178} & \cite{179} & \cite{180} \\ \hline 
 Secure cryptographic hash function \cite{155} &  v &  v & \checkmark  &  v &   v &  v &  v &  v &  v & \checkmark & \checkmark & \checkmark & \checkmark &  \\ \hline 
Chebyshev Chaotic Maps \cite{193} &  v &   &   &   &   &   &   &   &   &  &  &  &  &  \\ \hline 
 Chebyshev Polynomials \cite{194} &  v &   &   &   &   &   &   &   &   &  &  &  &  &  \\ \hline 
 ID-based cryptography (IBC) \cite{162} &   &  v &   &  v &   v &   &   &   &   &  &  &  &  &  \\ \hline 
 Advanced Encryption Standard (AES) \cite{200} &   &   &  v &   &   &   &   &   &   &  &  &  &  &  \\ \hline 
 Biometric &   &   &   &   &   & \checkmark &   &   &   &  &  &  &  &  \\ \hline 
 Password &   &   &   &   &   & \checkmark &   &   &   & \checkmark & \checkmark &  &  &  \\ \hline 
Smart card &   &   &   &   &   & \checkmark & \checkmark & \checkmark & \checkmark & \checkmark & \checkmark &  &  &  \\ \hline 
 Fuzzy extractor technique \cite{211} &   &   &   &   &   & \checkmark &   &   &   &  &  &  &  & \checkmark \\ \hline 
Elliptic Curve Diffie-Hellman (ECDH) \cite{85} &   &   &   &   &   &   & \checkmark & \checkmark &   &  &  &  &  &  \\ \hline 
 Key agreement &   &   &   &   &   &   &   &   & \checkmark & \checkmark & \checkmark &  &  &  \\ \hline 
Biohashing \cite{219} &   &   &   &   &   & \checkmark &   &   &   &  &  &  &  &  \\ \hline 
Access polynomial \cite{224} &  &  &  &  &  &  &  &  &  &  &  & \checkmark &  &  \\ \hline 
Elliptic curve cryptography \cite{225} &  &  &  &  &  &  &  &  &  &  &  &  & \checkmark & \checkmark \\ \hline 
\end{tabular}}
\label{tab:Tab4d}
\end{table}
\begin{table}[!]
\centering
\caption{The smart card-based authentication protocols}
{\tiny
\begin{tabular}{|p{1.2in}|p{1in}|p{3.8in}|} \hline 
\textbf{Protocol } & \textbf{Type} & \textbf{Design goal} \\ \hline 
Das (2016) \cite{172} & Remote authentication & - Providing a user authentication to resolve the security weaknesses of the scheme \cite{208}. \\ \hline 
Chang et al. (2016) \cite{173} & Remote authentication & - Providing mutual authentication and perfect forward secrecy. \\ \hline 
Jiang et al. (2016) \cite{174} & Remote authentication & - Providing mutual authentication, anonymity, and untraceability. \\ \hline 
Farash et al. (2016) \cite{175} & Remote authentication & - Providing the user authentication with traceability protection and sensor node anonymity. \\ \hline 
Srinivas et al. (2017) \cite{176} & Remote authentication & - Providing the mutual authentication with anonymity and unlinkability. \\ \hline 
\end{tabular}}
\label{tab:Tab4e}
\end{table}

As presented in Tab. \ref{tab:Tab4c}, the protocol \cite{127} uses two types of verification, namely, Heavy signing light verification (HSLV) and Light signing heavy verification (LSHV), which is based on the HORS scheme \cite{148}. The HSLV uses the following three algorithms\textit{: Key Generation}, \textit{Signing}, and\textit{ Verification}. The \textit{Key Generation} algorithm outputs the public key $PK=(k,v_1,v_2,\dots ,\ v_t)$  and the secret key $SK=(k,s_1,s_2,\dots ,\ s_t)$ where the trusted authority generates $t$ random $l$-bit strings $s_1,s_2,\dots ,\ s_t$. The signature is $(c,(\ s_{i1},s_{i2},\dots ,\ s_k))$ generated by the \textit{Signing} algorithm. To verify a signature  $(c',(\ {s'}_{i1},{s'}_{i2},\dots ,\ {s'}_k))$ over message $m$, the user check if he output integers  $i1>i2>ik$ and $f\left({s'}_j\right)=v_{ij}$ hold. On the other hand, with LSHV, the signature verification process verifies the $k$ elements of a signature by applying the one-way function for a distinct number of times over each element. Similar to the protocol \cite{92}, the protocol \cite{128} uses the same idea of Merkle Hash tree technique \cite{150}. In order to increase the level of security, the protocol \cite{129} uses three cryptosystems, namely, short signatures (BLS) \cite{152}, batch verification \cite{153}, signature aggregation \cite{154}. The BLS is introduced by Boneh-Lynn-Shacham \cite{152}, which is based on Gap Diffie--Hellman groups. Specifically, the BLS scheme uses the following three algorithms: 1) \textit{Key generation} algorithm to output the public key $v\in G_2$ and the private key $x$ , where $x\leftarrow Z_p$ and $v\leftarrow {g_2}^x$; 2) \textit{Signing} algorithm to generate a signature $\sigma \in G_1$, where $\sigma \leftarrow h^x$ and $h\leftarrow H(M)\in G_1$; and 3) \textit{Verification} algorithm to verify that $(g_2,v,h,\sigma )$ is a valid co-Diffie--Hellman tuple. The author of short signatures (BLS) \cite{152}, i.e., Dan Boneh, proposes the idea of signature aggregation \cite{154}, where an aggregate signature is valid only if it is an aggregation of signatures on distinct messages. Similar to the protocol \cite{89}, the protocol \cite{130} uses the same cryptosystem, i.e., Identity-based Public Key Cryptosystem \cite{111}. Therefore, both protocols \cite{131} and \cite{134} use the two same cryptosystems, namely, 1) the public-key encryption, such as RSA \cite{156}, and 2) HMAC, such as SHA-1 \cite{157} and MD5 \cite{158}. The protocol \cite{132} uses the Diffie-Hellman key establishment protocol \cite{159} in order to provide forward secrecy in Transport Layer Security's ephemeral modes. The protocol \cite{133} uses the EIBC mechanism \cite{161}, which is based on the original model developed by D. Boneh and M. Franklin. In addition, the protocol \cite{134} uses the Homomorphic Encryption \cite{164} and the Bloom Filter \cite{165}. The protocol \cite{135} uses two cryptosystems, 1) HMAC, such as SHA-1 \cite{157} and MD5 \cite{158}, 2) a symmetric encryption/decryption algorithm \cite{159}.

As presented in Tab.\ref{tab:Tab4d}, the protocol \cite{167} uses two countermeasures, namely, Chebyshev Chaotic Maps \cite{193} and Semigroup Property of Chebyshev Polynomials \cite{194}. The Chebyshev Polynomial of degree $p$ is defined by Mason and Handscomb \cite{194} as $T_p\left(x\right)={\rm cos}(p\ X\ {\rm acrcos}\ x)$ where the domain is the interval $x\in [-1,1]$ with two properties \cite{244}. However, three protocols, i.e., \cite{168}, \cite{170}, and \cite{171} use the ID-based cryptography (IBC) \cite{162}. On the other hand, the protocol \cite{169} uses the Advanced Encryption Standard (AES) \cite{200} such as the protocol \cite{117}. The smart card-based authentication protocols is a very promising and practical solution to remote authentication \cite{245}, as presented in Tab. \ref{tab:Tab4e}. There are five \cite{172,173,174,175,176} smart card-based authentication protocols where each protocol integrates a method with the smart card. For example, the protocol \cite{172} uses the fuzzy extractor technique \cite{211}, which a fuzzy extractor is a pair of randomized procedures, ``generate'' (Gen) and ``reproduce'' (Rep), and is efficient if Gen and Rep run in expected polynomial time. For more details about the fuzzy extractor technique, we refer the reader to the paper \cite{211}. In addition, the elliptic curve cryptography \cite{225} is used by both protocols \cite{179} and \cite{180}.\textbf{}

\begin{figure}[h]
            \centering
            \includegraphics[width=0.3\linewidth]{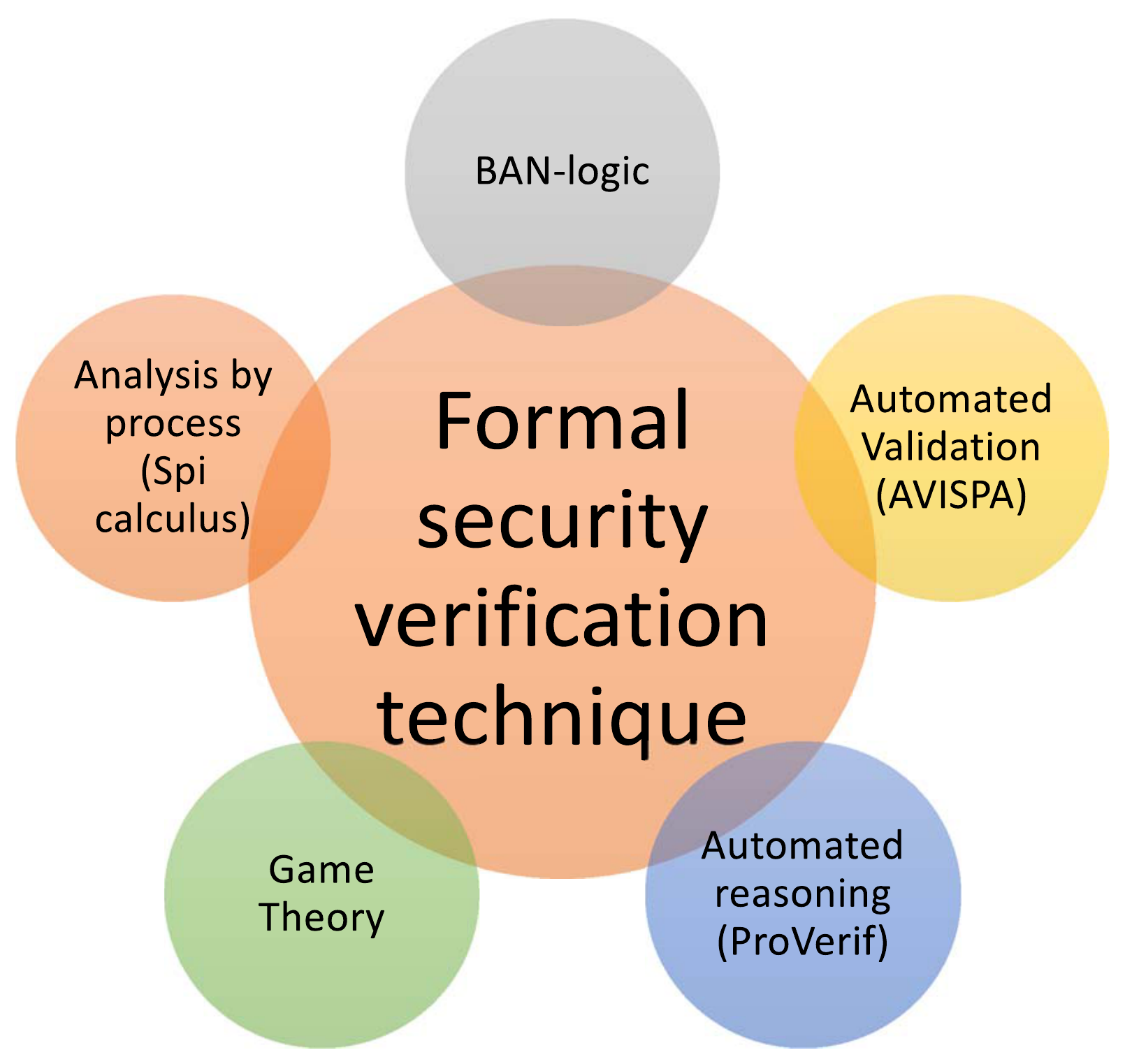}
            \caption{Formal security verification techniques}
            \label{fig:Fig4b}
            \end{figure}

\begin{table}[!]
\centering
\caption{Formal security verification techniques used in Authentication protocols for the IoT}
{\tiny
\begin{tabular}{|p{0.8in}|p{3in}|p{3in}|} \hline 
\textbf{Protocol} & \textbf{Approach} & \textbf{Main results} \\ \hline 
Lai et al. (2013) \cite{79} & - The security of the protocol is analyzed using the ProVerif tool \cite{83} & - Proof the mutual authentication between mobile equipment and its serving network. \\ \hline 
Shao et al. (2016) \cite{91} & - Decisional Diffie-Hellman (DDH) Assumption;\newline - Decision Linear (DLIN) Assumption;\newline - Extended Computational Diffie-Hellman (eCDH) Assumption\newline - Computational Inverse Diffie-Hellman (ciCDH) Assumption & - The proposed group signature scheme satisfies unforgeability;\newline The proposed group signature scheme satisfies anonymity;\newline - The proposed theorem satisfies the traceability. \\ \hline 
Zhang et al. (2016) \cite{93} & - Based on the size of the beacon interval and the network bandwidth. & - Broadcasting the MAC of a message's prediction outcome is secure. \\ \hline 
Zhang et al. (2015) \cite{108} & - Bilinear Diffie-Hellman and the computational Diffie- Hellman assumptions & - The protocol satisfies individual authentication, non-repudiation, vehicle privacy and traceability. \\ \hline 
Dolev et al. (2016) \cite{94} & - Spi calculus \cite{115} & - The proposed session key establishment protocol respects the authenticity property and the secrecy property. \\ \hline 
Chan et al. (2014) \cite{95} & - NXP-ATOP platform \cite{116} & - Demonstrate the two-factor cyber-physical device authentication \\ \hline 
Lai et al. (2013) \cite{136} & - The security of the protocol is analyzed using the ProVerif tool \cite{83} & - The scheme can implement mutual authentication and key agreement between multiple devices and the core network simultaneously. \\ \hline 
Li et al.  (2011) \cite{127} & - Prove the existence of a pivot rank by contradiction. & - The total signing cost does not increase.  \\ \hline 
Li et al. (2012) \cite{129} & - Diagnose tools & - Detect failure points and to minimize the whole fault time. \\ \hline 
Nicanfar et al. (2014) \cite{133} & - Automated Validation of Internet Security Protocols and Application (AVISPA) security analyzer \cite{163} & - Providing mutual authentication and key management mechanisms. \\ \hline 
Mahmood et al. (2016) \cite{135} & - The security of the protocol is analyzed using the ProVerif tool \cite{83} & - Verifies mutual authentication and session key secrecy properties of the proposed scheme. \\ \hline 
Kumari et al. (2016) \cite{167} & - Burrows-Abadi-Needham Logic (BAN-logic) \cite{192}. & - Prove that the proposed scheme establish a session key between user and sensor node \\ \hline 
Chung et al. (2016) \cite{168} & - Burrows-Abadi-Needham Logic (BAN-logic) \cite{192}. & - Prove the validity of authentication and key agreement protocol. \\ \hline 
Amin et al. (2016) \cite{170} & - Burrows-Abadi-Needham Logic (BAN-logic) \cite{192}.\newline - Automated Validation of Internet Security Protocols and Application (AVISPA) security analyzer \cite{163}. & - Prove that the protocol has achieved mutual authentication and session key agreement securely. \\ \hline 
Das (2016) \cite{172} & - Automated Validation of Internet Security Protocols and Application (AVISPA) security analyzer \cite{163}. & - The scheme is secure against the replay and man-in-the-middle attacks against an adversary. \\ \hline 
Chang et al. (2016) \cite{173} & - Sequence of games under the decisional Diffie-Hellman (ECDDH) problem. & - The scheme provides secure and perfect forward secrecy authentication. \\ \hline 
Jiang et al. (2016) \cite{174} & - Burrows-Abadi-Needham Logic (BAN-logic) \cite{192}. & - The improved scheme accomplishes mutual authentication and key agreement between the user and sensor, the user and the gateway node. \\ \hline 
Farash et al. (2016) \cite{175}\newline  & - Burrows-Abadi-Needham Logic (BAN-logic) \cite{192}.\newline - Automated Validation of Internet Security Protocols and Application (AVISPA) security analyzer \cite{163}. & - Prove that the scheme allows a user to establish a session key with a sensor node of his choice near the end of the authentication process. \\ \hline 
Srinivas et al. (2017) \cite{176} & - Burrows-Abadi-Needham Logic (BAN-logic) \cite{192}.\newline - Automated Validation of Internet Security Protocols and Application (AVISPA) security analyzer \cite{163}. & - The scheme can resist numerous security attacks, which include the attacks, found in the Amin and Biswas's scheme \cite{170}. \\ \hline 
Kumari et al. (2016) \cite{177} & - Burrows-Abadi-Needham Logic (BAN-logic) \cite{192}. & - The scheme provides secure mutual authentication between a legal user and an accessed sensor node inside WSN or not. \\ \hline 
Jiang et al. (2016) \cite{179}\newline  & - Burrows-Abadi-Needham Logic (BAN-logic) \cite{192}. & - Prove that an identity and a session key is agreed between the user and the sensor \\ \hline 
Wu et al. (2016) \cite{180} & - The security of the protocol is analyzed using the ProVerif tool  \cite{83} & - The scheme passes the verifications according to the Dolev-Yao model \cite{226}. \\ \hline 
Das et al. (2016) \cite{181} & - Burrows-Abadi-Needham Logic (BAN-logic) \cite{192}.\newline - Random oracle model.\newline - Automated Validation of Internet Security Protocols and Application (AVISPA) security analyzer \cite{163}. & - Prove secure mutual authentication between a legal user and an accessed sensor node. \\ \hline 
Das et al. (2016) \cite{184} & - Automated Validation of Internet Security Protocols and Application (AVISPA) security analyzer \cite{163}. & - The scheme is free from man-in-the-middle and replay attacks. \\ \hline 
\end{tabular}
}
\label{tab:Tab4f}
\end{table}

\subsection{Formal security verification techniques}

In order to prove the performance of an authentication protocol in terms of security, researchers use formal security verification techniques. As presented in Fig. \ref{fig:Fig4b}, there are five formal security verification techniques, namely, BAN-logic, analysis by process (Spi calculus), Game Theory, Automated reasoning (ProVerif), and Automated Validation (AVISPA). In addition, Tab. \ref{tab:Tab4f} presents the formal security verification techniques used in authentication protocols for the IoT.

The Burrows-Abadi-Needham Logic (BAN-logic) \cite{192} is used by nine authentication protocols \cite{167,168,170,174,175,176,177,179,181}. A typical BAN logic sequence includes three steps, 1) Verification of message origin; 2) Verification of message freshness; and 3) Verification of the origin's trustworthiness. Therefore, the protocol \cite{167} uses the BAN-logic to prove that the proposed protocol can establish a session key between user and sensor node. Both protocols \cite{168} and \cite{179} use the BAN-logic in order to prove that the protocol has achieved mutual authentication and session key agreement securely. The protocol \cite{176} uses the BAN-logic to prove that the protocol can resist numerous security attacks, which include the attacks, found in the Amin and Biswas's scheme \cite{170}. There are seven authentication protocols \cite{133,170,172,175,176,181,184} that use the Automated Validation of Internet Security Protocols and Application (AVISPA) security analyzer \cite{163}. The AVISPA tool provides a modular and expressive formal language for specifying security protocols and properties. The protocol \cite{184} uses the AVISPA tool in order to prove that the proposed protocol is free from man-in-the-middle and replay attacks. The protocol \cite{175} uses the AVISPA tool to prove that the protocol allows a user to establish a session key with a sensor node of his choice near the end of the authentication process. In addition, there are four authentication protocols \cite{79,136,135,180} that use the ProVerif tool \cite{83}, which is an automatic cryptographic protocol verifier, in the formal model, called Dolev-Yao model \cite{226}. The protocol \cite{79} uses the ProVerif tool in order to proof the mutual authentication between the mobile equipment and its serving network. The protocol \cite{136} uses the ProVerif tool to prove that the proposed protocol can implement mutual authentication and key agreement between multiple devices and the core network simultaneously. The protocol \cite{180} uses the ProVerif tool prove that the proposed protocol can pass the verifications according to the Dolev-Yao model \cite{226}. Finally, the protocol \cite{173} uses a sequence of games under the decisional Diffie-Hellman (ECDDH) problem in order to proof that the protocol provides secure and perfect forward secrecy authentication. For more details about the game-theoretic approaches, we refer the reader to the survey \cite{246}.
 
\begin{figure}[h]
            \centering
            \includegraphics[width=0.6\linewidth]{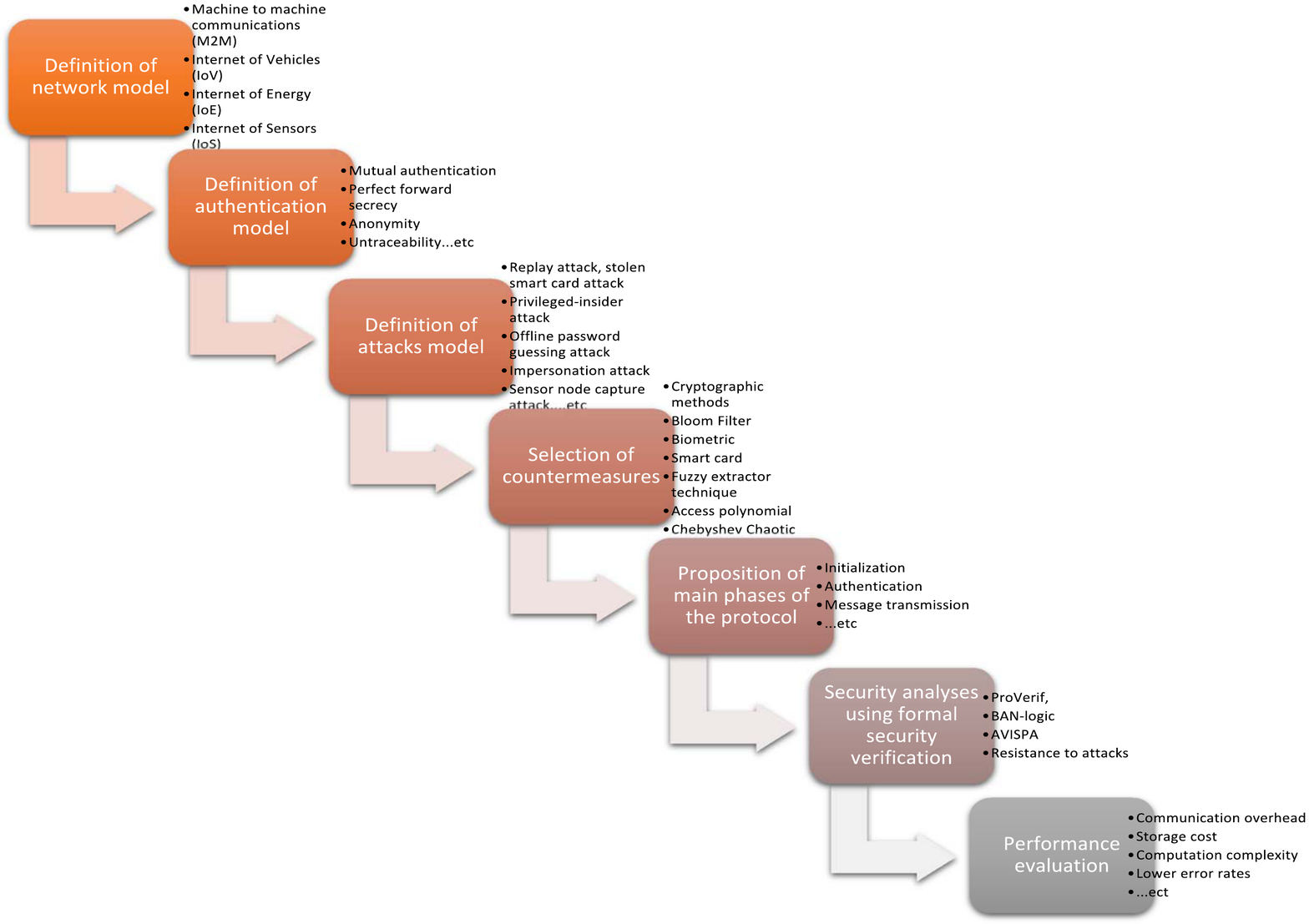}
            \caption{The realization processes of an authentication protocol for the IoT}
            \label{fig:Fig5a}
            \end{figure}
            
\section{Taxonomy and comparison of authentication protocols for the IoT}\label{sec:taxonomy-and-comparison-of-authentication-protocols-for-the-iot}
In this section, we in-detail examine authentication protocols developed for or applied in the context of IoT. The detailed comparison of computational cost and communication overhead of authentication protocols for the IoT and the notations used are given in Tab. 19 and \ref{tab:Tab5a}, respectively. As shown in Fig. \ref{fig:Fig5a}, the realization processes of an authentication protocol for IoT are based on the following processes: 

\begin{enumerate}
\item  Definition of network model (e.g., M2M, IoV, IoE, IoS...etc.).
\item  Definition of authentication model (e.g., mutual authentication, perfect forward secrecy, anonymity, untraceability...etc.).
\item  Definition of attacks model (e.g., replay attack, stolen smart card attack, privileged-insider attack, offline password guessing attack, impersonation attack, and sensor node capture attack....etc.).
\item  Selection of countermeasures (e.g., cryptographic methods, bloom Filter, biometric, Smart card, access polynomial, Chebyshev Chaotic Maps \dots etc.).
\item  Proposition of main phases of the protocol (e.g., initial setup; registration process...etc.).
\item  Security analyses using formal security verification (e.g., ProVerif, BAN-logic, AVISPA\dots etc).
\item  Performance evaluation (e.g., in terms of storage cost, computation complexity, communication overhead, lower error rates\dots etc.).
\end{enumerate}

\begin{table}[!]
\centering
\caption{Notations used in comparison of computational cost and communication overhead}
{\tiny
\begin{tabular}{|p{6in}|} \hline 
\textbf{\underbar{Notations}}\newline 
$n$: represents the number of machine-type communications (MTC) devices in a group; \newline $m$ : represents the number of groups;\newline ${T\ }_{EAP}$ : The latency for the full Extensible Authentication Protocol authentication;\newline $T_w$: The transmission latency between the Mobile Station and Base Station;\newline $T_c$: The transmission latency between any two relatively close devices;\newline $T_a$: The transmission latency between the Base Station and infrastructures;\newline $T_{mul}$: The time for a point multiplication operation;\newline $T_s$: The time for a symmetric encryption or decryption operation;\newline $T_{MAC}$: The time for a HMAC operation;\newline $T_{pair}$: The time for a bilinear pairing operation;\newline $T_D$: The time for a Dot16KDF operation;\newline $T_{hash}$: The time for a hash operation;\newline $T_{mtp}$: The time for a map to point hash operation;\newline $E_{handshake}$: The energy consumption in audio handshake phase;\newline $E_{audiorec}$: The energy consumption during Sound(AuthA) reception;\newline $E_{feature}$: The energy consumption during the fingerprint extraction phase;\newline $E_{MA}$: The energy consumption during the fingerprint matching phase;\newline $E_{audiogen}$: The energy consumption during the audio signals Sound(AuthA) generation;\newline $E_{audiotran}$: The energy consumption during the audio signals Sound(AuthA) transmission;\newline $E_{dete}$: The energy consumption in the active attack detection algorithm;\newline $M:$ Message length(bit);\newline $\alpha $: The transmission cost of an authentication message between the mobile subscribers and the visiting authentication server;\newline $\beta $: The transmission cost of an authentication message between the visiting authentication server and the home authentication center;\newline $T_e$: The computation costs of an exponentiation operation in $G_1$;\newline $T_{enc}$: The time of symmetric key encryption operations;\newline $B_{FMAGs\_list}$: The transmitted bytes of Mobile Access Gateway;\newline $B_{CHL/RESP}$ : The transmitted bytes of the challenge/response messages during handovers;\newline $T_{disclose}$: The require time for disclose;\newline $B_{ACK}$: The transmitted bytes of Ack messages;\newline $B_{disclose}$: The transmitted bytes of disclose;\newline $T_{EAP}$: The require time for the Extensible Authentication Protocol;\newline $B_{EAP}$: The transmitted bytes for the Extensible Authentication Protocol;\newline $B_{key-exchange}$: The transmitted bytes of key exchange phase;\newline $T_{ver}$: The require time for verifying signatures;\newline $T_{sing}$: The require time for signing;\newline $B_{Cert}$: The transmitted bytes of a certificate;\newline $N_{rev}$: The total number of revoked certificates in a certificate revocation list;\newline $t_{e,G_1}$: The timings for one exponentiation in G1;\newline $t_{e,G_2}$: The timings for one exponentiation in G2;\newline $T_{ASED}$: The time for asymmetric encryption/decryption;\newline $T_{SED}$: The time for symmetric encryption/decryption;\newline $T_{HMAC}$: The time for HMAC;\newline $T_{Chaotic}$: The time complexity for computing Chebyshev chaotic map operation;\newline $T_{ECP}$: The time complexity for computing elliptic curve point multiplication;\newline $T_{mod}$:The execution time of a modular exponential operation;\newline $T_{fuzzy}$: The time for the fuzzy extractor operation;\newline $T_{XOR}$: The time for performing an XOR operation; \\ \hline 
\end{tabular}}
\label{tab:Tab5a}
\end{table}
\clearpage
\begin{center}
\topcaption{Comparison of computational cost and communication overhead}
\end{center}
{\tiny
\begin{supertabular}{|p{1in}|p{2.7in}|p{2in}|} \hline 
\textbf{Protocol} & \textbf{Computation complexity} & \textbf{Communication overhead} \\ \hline 
Lai et al. \cite{76} & GLARM-1: $(4T_{hash})n+3T_{hash}$\newline GLARM-2 : $(4T_{hash})n+3T_{hash}$ & N/A \\ \hline 
Cao et al. \cite{77} & $nT_{mtp}+\ (2n+1)T_{mul}+2T_{pair}$ & $2n+6m$ \\ \hline 
Lai et al. \cite{136} & $(3\ +\ 2n)T_{hash}$ & $6m$ \\ \hline 
EPS \cite{146}  & $(5T_{hash})n$ & $8n$ \\ \hline 
EAP-AKA  & $(5T_{hash})n$ & N/A \\ \hline 
Zhang et al. \cite{249} & $(4T_{hash})n$ & N/A \\ \hline 
Huang et al. \cite{250} & $(6T_{hash})n$ & N/A \\ \hline 
Chen et al. \cite{251} & $(3T_{hash})n+(2T_{hash})m$ & N/A \\ \hline 
Chen et al. \cite{71} & $2E_{handshake}+2E_{audiorec}+E_{feature}+E_{MA}+E_{audiogen}+E_{audiotran}+E_{dete}$ & N/A \\ \hline 
Lai et al. \cite{78} & $nT_{mtp}+(2n+\ 1)T_{mul}+2T_{pair}$ & N/A \\ \hline 
Lai et al. \cite{79} & $(3T_{hash}+2T_{mul})n+(2T_{hash})m$ & N/A \\ \hline 
Fu et al. \cite{81} & $2T_{MAC}+{\ T}_D+\ 2T_{mul}$ & $2T_c/n\ +\ 4T_w$ \\ \hline 
Huang et al. \cite{253} & $2T_s+2T_{MAC}+T_D$ & $2T_c\ +\ 5T_w$ \\ \hline 
Zhang et al. \cite{254} & $2T_s+2T_{MAC}+2T_{pair}+3T_{mul}$ & $2T_c\ +\ 5T_w$ \\ \hline 
Lai et al. \cite{118} & $25.25T_e+10T_{pair}$ & $3\alpha $ \\ \hline 
Yang et al. \cite{121} & $8.75T_e+3T_{pair}$ & $3\alpha $ \\ \hline 
He et al. \cite{122} & $15.75T_e+4T_{pair}$ & $3\alpha $ \\ \hline 
Kim et al. \cite{255} & N/A & $3\alpha $ \\ \hline 
Jiang et al. \cite{256} & N/A & $3\alpha +2\beta $ \\ \hline 
Shi et al. \cite{257} & N/A & $4\alpha +2\beta $ \\ \hline 
Céspedes et al. \cite{89} & $2T_{enc}$ & $B_{FMAGs\_list}\ +\ B_{CHL/RESP}$ \\ \hline 
Xie et al. \cite{259} & $T_{enc}$+$T_{ver}$+$T_{sing}$ & $B_{Cert}$ \\ \hline 
Ristanovic et al. \cite{260} & $T_{sing}+T_{ver}$ & $B_{Cert}$ \\ \hline 
Wasef et al. \cite{90} & $O(N_{rev})$ & The revocation check=$0.42\ \mu sec$ \\ \hline 
Shao et al. \cite{91} & ${15t}_{e,G_1}+{5t}_{e,G_2}+{13t}_{pair}$ & N/A \\ \hline 
Lyu et al. \cite{92} & N/A & N/A \\ \hline 
Al Shidhani \cite{258}  & $T_{enc}+T_{EAP}$ & $B_{EAP}+B_{key-exchange}$ \\ \hline 
Zhang et al. \cite{93} & N/A & N/A \\ \hline 
Heer et al. \cite{247} & $T_{enc}$+$T_{disclose}$ & $B_{ACK}+B_{disclose}$ \\ \hline 
Zhang et al. \cite{108} & N/A & N/A \\ \hline 
Dolev et al. \cite{94} & Direct iteration cost = $2\ rounds$ & N/A \\ \hline 
Krawczyk \cite{113} & Direct iteration cost = $3\ rounds$ & N/A \\ \hline 
Chan et al. \cite{95} & N/A & N/A \\ \hline 
Sun et al. \cite{117} & $(4T_{hash})n$ & User to M2M server : $M=384$\newline M2M server to User : $M=160$ \\ \hline 
Lai et al. \cite{125} & $\left(4T_{hash}\right)n+(7n+5)T_{mul}$ & N/A \\ \hline 
Li et al.  \cite{127} & Signing \& verification cost =$\mu +\frac{k(k+1)}{2}+1$ & Signing size (bit)= $kl+log\mu $ \\ \hline 
Perrig \cite{261} & Signing \& verification cost $=2t+2k+1$ & Signing size (bit)= $kl$ \\ \hline 
Li et al. \cite{128} & With the number of HAN users = 4000, the computation cost = $0.36\ ms$ & $896\ bits$ \\ \hline 
IEEE 802.16m \cite{252} & $3T_s+6T_{MAC}+T_D$ & ${T\ }_{EAP}+\ T_a\ +\ 5T_w$ \\ \hline 
Chim et al. \cite{134} & $38.2n\ +\ 1390.8\ msec$ & N/A \\ \hline 
Mahmood et al. \cite{135} & $6T_{ASED}+4T_{SED}+2T_{hash}+2T_{HMAC}$ & $800\ bytes$ \\ \hline 
Fouda et al. \cite{140} & $9T_{ASED}+2T_{hash}$ & $1040\ bytes$ \\ \hline 
Sule et al. \cite{166} & $9T_{ASED}+2T_{hash}+2T_{HMAC}$ & $1040\ bytes$ \\ \hline 
Kumari et al. \cite{167} & ${13T}_{hash}+4T_{Chaotic}+4T_{SED}$ & $1408\ bits$ \\ \hline 
Xue et al. \cite{203} & ${22T}_{hash}$ & $2432\ bits$ \\ \hline 
Li et al. \cite{195} & ${28T}_{hash}$ & $2432\ bits$ \\ \hline 
Li et al. \cite{262} & ${18T}_{hash}+10T_{SED}$ & $1152\ bits$ \\ \hline 
He et al. \cite{196} & ${23T}_{hash}$ & $1920\ bits$ \\ \hline 
Choi et al. \cite{213} & ${{6T}_{ECP}+20T}_{hash}$ & $2944\ bits$ \\ \hline 
Chung et al. \cite{168} & ${28T}_{hash}$ & N/A \\ \hline 
Jiang et al. \cite{263} & ${15T}_{hash}+2T_{mod}$ & N/A \\ \hline 
Wen et al. \cite{264} & ${15T}_{hash}+4T_{mod}$ & N/A \\ \hline 
Shin et al. \cite{265} & ${14T}_{hash}+4T_{SED}+2T_{mod}$ & N/A \\ \hline 
Gope et al. \cite{266} & ${18T}_{hash}+2T_{SED}+2T_{mod}$ & N/A \\ \hline 
Farash et al. \cite{197} & ${23T}_{hash}+4T_{SED}$ & N/A \\ \hline 
Amin et al. \cite{170} & ${20T}_{hash}$ & $1280/1680\ bits$ \\ \hline 
Turkanoviæ et al. \cite{202} & ${19T}_{hash}$ & $768\ bits$ \\ \hline 
Das et al. \cite{204} & ${10T}_{hash}+4T_{SED}$ & $384\ bits$ \\ \hline 
Turkanoviæ and Hölbl \cite{205} & ${7T}_{hash}+5T_{SED}$ & $384\ bits$ \\ \hline 
Yeh et al. \cite{206} & ${8T}_{hash}+8T_{SED}$ & $384\ bits$ \\ \hline 
Chen and Shih \cite{216} & ${10T}_{hash}$ & $384\ bits$ \\ \hline 
Fan et al. \cite{268} & ${19T}_{hash}$ & $640\ bits$ \\ \hline 
Vaidya et al. \cite{215} & ${13T}_{hash}$ & $384bits$ \\ \hline 
He et al. \cite{269} & ${11T}_{hash}$ & $384\ bits$ \\ \hline 
Huang et al. \cite{270} & ${11T}_{hash}$ & $384\ bits$ \\ \hline 
Gope et al. \cite{171} & ${19T}_{hash}$ & $35\ bytes$ \\ \hline 
Das \cite{172} & $31T_{hash}\ +\ 2T_{fuzzy}$ & $4\ messages\ (1952\ bits)$ \\ \hline 
Chang et al. \cite{173} & ${20T}_{hash}+9T_{XOR}$ & N/A \\ \hline 
Khan and Alghathbar \cite{267} & ${12T}_{hash}$ & $384\ bits$ \\ \hline 
Jiang et al. \cite{174} & ${23T}_{hash}+3T_{ECP}$ & N/A \\ \hline 
Farash et al. \cite{175} & $32T_{hash}$ & User - Sensor node $=\ 79\ bytes$ \newline
Sensor node - Gateway node $=\ 158\ bytes$ \newline
Gateway node - Sensor node $=\ 99\ bytes$ \newline
Sensor node - User $=\ 98\ bytes$
\\ \hline 
Srinivas et al. \cite{176} & Case-1: $29T_{hash}$\newline Case-2: $35T_{hash}$ & Case-1 = $4\ messages\ (353\ bytes)$\newline Case-2 = $7\ messages\ (547\ bytes)$ \\ \hline 
Kumari et al. \cite{177} & $44T_{hash}$ & $58\ bytes$ \\ \hline 
Jiang et al. \cite{179} & $17T_{hash}+6T_{ECP}$ & N/A \\ \hline 
\end{supertabular}}
\label{tab:Tab5b}
\begin{center}
\topcaption{Summary of authentication protocols for M2M (Published between 2012 and 2016)}
\end{center}
{\tiny
\begin{supertabular} {|p{0.5in}|p{1in}|p{1in}|p{1in}|p{3in}|} \hline 
\textbf{Protocol } & \textbf{Network model} & \textbf{Goals} & \textbf{Main processes} & \textbf{Performances (+) and limitations (-)} \\ \hline 
Lai et al. (2016) \cite{76} & - Based on 3GPP standard with three domains, including, access networks, evolved packet core, and non-3GPP domain, e.g., Internet. & - Guarantee the entity mutual authentication and secure key agreement. & - Initialization phase;\newline - Group authentication and key agreement phase. & + Resistance to DoS attack, redirection attack, and man-in-the-middle attack.\newline + Computation overheads are fairly small.\newline + Computation complexity is much less than schemes \cite{77} \cite{78} and \cite{79}.\newline + Can ensure QoS for machine-type communications devices.\newline - Some privacy models are not analyzed such as location privacy and identity privacy.\newline - Storage costs is not considered. \\ \hline 
Chen et al. (2016) \cite{71} & - Two wireless devices. & - Achieving variable distance authentication and active attack detection. & - Audio-handshake phase; \newline - Mixed-signal generation phase;\newline - Feature extraction and storage phase. & + Efficient in terms of lower error rates compared with DISWN \cite{72}, LDTLS \cite{73}, PLTEA \cite{74}, and SeArray \cite{75}.\newline + Active attack detection (e.g., audio replay attack).\newline - Privacy-preserving is not analyzed compared to the GLARM scheme \cite{76}.\newline - Storage costs is not considered. \\ \hline 
Lai et al. (2014) \cite{78}\newline \newline \newline  & - 3GPP-WiMAX-Machine-type Communication. & - Achieving mutual authentication and key agreement between all Machine-type Communication devices. & - Initialization phase\newline - Roaming phase & + Efficient in terms of the communication overhead compared to the traditional roaming authentication scheme and the optimized roaming authentication scheme in \cite{81}.\newline + Efficient in terms of computation complexity compared to the scheme without aggregation.\newline - Resistance to attacks is not studied.\newline - Privacy-preserving is not analyzed compared to the GLARM scheme \cite{76}.\newline - Storage costs is not considered.\newline  \\ \hline 
Lai et al. (2013) \cite{79}\newline \newline  & - 3GPP standard with three domains, namely access network domain, serving network domain and home network domain. & - Guarantee privacy-preservation and key forward/backward secrecy with. & - Preparation and initialization;\newline - Protocol execution for the first equipment;\newline - Protocol execution for the remaining equipment of the same group;\newline - Group member joining/leaving the group. & + Considers the data integrity and ensure user privacy.\newline + Resistance to attacks (DoS attack, redirection attack, man-in-the-middle attack, and replay attack) \newline + The overhead of authentication message delivery of SE-AKA is lower than other existing AKA protocols.\newline + The computational overhead is larger than that of other traditional protocols such as the work \cite{84}.\newline + Smaller storage costs than others protocols\newline - Some privacy models are not analyzed such as location privacy and identity privacy. \\ \hline 
Fu et al. (2012) \cite{81}\newline  & - Mobile WiMAX networks with an access service network & - Achieving mutual authentication and privacy preservation, and  resisting the domino effect. & - Pre-deployment phase;\newline - Initial authentication phase;\newline - Handover authentication phase. & + Efficient in terms of the computational and communication overhead compared to three schemes \cite{87} \cite{88} \cite{89}.\newline + Considers the privacy preservation.\newline - Storage costs is not considered.\newline - Resistance to attacks is not studied.\newline - No threat model presented.\newline - Error-detection and fault tolerance\newline are not considered. \\ \hline 
Sun et al. (2015) \cite{117} & - Mobile users, home gateways, and an M2M server. & - Achieving a mutual authentication process in machine-to machine home network service. & - Set-up;\newline - Registration phase;\newline - Login and authentication phase;\newline - Update password phase;\newline - Home gateway joins the Time Division-Synchronous Code Division Multiple Access network. & + Efficient in terms of the amount of calculation and communication volume compared to the protocol in \cite{119}.\newline + Resistance to guessing attack, stolen-verifier attack, impersonation attack, and replay attack.\newline - Privacy-preserving is not analyzed compared to the GLARM scheme \cite{76}.\newline - Storage costs is not considered.\newline - Lack non-repudiation compared to the PBA scheme in \cite{92}. \\ \hline 
Lai et al. (2014) \cite{118} & - Roaming network architecture with the home authentication center (HAC), the trust linking server (TLS), and the visiting authentication server (VAS). & - Providing a strong anonymous access authentication.\newline - Guarantee user tracking on a disputed access request.\newline - Achieving anonymous user linking and efficient user revocation for dynamic membership. & - System initialization;\newline - Roaming;\newline - User tracking algorithm;\newline - Anonymous user linking;\newline - User revocation. & + Efficient in terms of communication overhead and computation cost compared to two strong anonymous schemes \cite{121} and \cite{122}.\newline + Considers the data integrity and ensure user privacy.\newline + Resistance to attacks, namely, Denial of Service (DoS) attack and impersonation attack.\newline - Some privacy models are not analyzed such as location privacy.\newline - Lack non-repudiation compared to the PBA scheme in \cite{92}. \\ \hline 
Zhu et al. (2015) \cite{124} & - Android smartphone devices. & - Satisfy the user-friendly with a reasonable false rejection rate.\newline - Achieving an authentication process for Android smartphone devices.\newline  & - Feature-set extraction and storing for registration;\newline - Dual-factor authentication. & + Can enhance user-friendliness. \newline + Improve security without adding extra hardware devices.\newline - No threat model presented. \\ \hline 
Lai et al. (2013) \cite{136} & - Based on 3GPP standard with three domains: (1) Evolved universal terrestrial radio access network (E UTRAN), (2) Evolved Packet Core (EPC), (3) Non-3GPP domain, e.g., Internet. & - Guarantee the entity mutual authentication and secure key agreement. & - Initialization phase;\newline - Group authentication and key agreement phase. & + Efficient in terms of the signaling and computation overhead compared to the schemes \cite{77} \cite{146}.\newline + Resistance to attacks, i.e., replay attack, redirection attack, and man-in-the-middle attack.\newline + Considers the data integrity.\newline - Privacy-preserving is not analyzed compared to the GLARM scheme \cite{76}.\newline - Storage costs is not considered.\newline - Lack non-repudiation compared to the PBA scheme in \cite{92}. \\ \hline 
\end{supertabular}}
\label{tab:Tab5c}

\begin{center}
\topcaption{Summary of authentication protocols for IoE (Published between 2011 and 2016)}
\end{center}
{\tiny
\begin{supertabular} {|p{0.5in}|p{1in}|p{1in}|p{1in}|p{3in}|} \hline 
\textbf{Protocol } & \textbf{Network model} & \textbf{Goals} & \textbf{Main processes} & \textbf{Performances (+) and limitations (-)} \\ \hline 
Li et al. \newline (2011)\newline \cite{127}\newline  & - Smart Grid with wide multicast applications, namely, wide area protection, demand-response, operation and control, and in-substation protection. & - Provide multicast authentication. & - Key generation;\newline - Signing;\newline - Verification. & + Efficient in terms of hash or one-way function invocations compared to the scheme \cite{147}.\newline + Resistance to message forgery attacks.\newline + Can reduces the storage cost.\newline - Privacy-preserving is not discussed.\newline - The reports' confidentiality and integrity are not considered compared to the scheme \cite{128}. \\ \hline 
Li et al. (2014) \cite{128}\newline  & - Communication between the home area networks (HANs) and the neighborhood gateway using WiFi technology. & - Detecting the replay attacks;\newline - Providing authentication for the source of electricity consumption reports;\newline - Guarantees the reports' confidentiality and integrity. & - System initialization;\newline - Report generation;\newline - Neighborhood gateway authentication. & + Efficient in terms of  computation complexity of the HAN user and the neighborhood gateway compared to the RSA-based authentication scheme.\newline + Efficient in terms of communication overhead between the HAN user and the neighborhood gateway compared to the RSA-based authentication scheme.\newline + Resistance to attacks, namely, replay attack, message injection attack, message analysis attack, and message modification attack.\newline + Guarantees the reports' confidentiality and integrity compared to the scheme \cite{127}.\newline - The routing attacks are not considered such as wormhole attack. \\ \hline 
Li et al. (2012) \cite{129} & - The smart grid with power generation, power transmission, and power distribution. & - Providing the authentication for power usage data aggregation in Neighborhood Area Network (NAN) with fault tolerance architecture. & - Key generation;\newline - Signature generation;\newline - Batch verification and trinary diagnose TreeBatch;\newline - Signature amortization for Package Blocks; & + Makes significant performance gains in terms of the communication and computation cost.\newline + Considers the fault diagnosis.\newline - No threat model presented. \\ \hline 
Nicanfar et al. (2011) \cite{130}\newline  & - The data communication in outside of the Home Area Network (HAN).\newline - Some smart meters and a utility server under a wireless mesh network topology.\newline - Communication inside a home domain is IEEE 802.15.4 based, and outside is WiMax (IEEE 802.16).\newline - All node have an IP address (most likely IPv6). & Providing mutual authentication scheme to prevent brute-force attacks, replay attacks, Man-In-The-Middle (MITM) attack, and Denial-of-Service (DoS) attacks.  & - Initialization;\newline - Ongoing maintenance or Short period key\newline refreshment;\newline - Long period key refreshment;\newline - Multicast key support. & + Can provide simplicity and low overhead.\newline + Resistance to attacks, namely, brute-force attacks, replay attacks, Man-In-The-Middle (MITM) attack, and Denial-of-Service (DoS) attacks.\newline + Can provide secure key management.\newline - The reports' confidentiality and integrity are considered compared to the scheme \cite{128}. \\ \hline 
Chim et al. (2011) \cite{131}\newline  & - Smart grid network with three basic layers, namely, power generators, substations, and smart meters and smart appliances. & - Guarantee the message authentication, identity privacy, and traceability. & - Preparation module;\newline - Pseudo-identity generation module;\newline - Signing module;\newline - Verification module;\newline - Tracing module. & + Requires only an additional 368 msec for HMAC signature verification at a substation.\newline + Efficient in overall normal traffic success rate when under attack.\newline + The message overhead is only 20 bytes per request message.\newline - The routing attacks are not considered such as wormhole attack.\newline - Storage costs is not considered.\newline - No comparison with other schemes. \\ \hline 
Fouda et al. (2011) \cite{132} & - Smart grid with the power Distribution Network (DN), the Transmission Substation (TS), and a number of Distribution Substations (DSs).\newline - The communication framework for the lower distribution network split into NAN, BAN, and HAN.\textbf{} & - Providing mutual authentication and achieve message authentication in a light-weight way. & - Key generation;\newline - Message generation;\newline - Hash-based message authentication. & + Efficient in terms of communication overhead and message decryption/verification delay compared to ECDSA-256.\newline + Resistance to attacks, namely, replay attack, chosen-plaintext attack, and collision attack.\newline - Location privacy is not considered.\newline - Identity privacy and traceability are not considered compared to the scheme \cite{131}. \\ \hline 
Nicanfar et al. (2014) \cite{133} & - Multigate communication network proposed in ref. \cite{160} & - Providing mutual authentication and key management mechanisms.  & - SGMA scheme (System setup; Mutual authentication Scheme)\newline - SGKM protocol (Key refreshment; Multicast key mechanism; Broadcast key mechanism) & + Can prevent the adversary from continuing the successful attack.\newline + Can prevent various attacks while reducing the management overhead.\newline - Storage costs is not considered.\newline - Lack non-repudiation compared to the PBA scheme in \cite{92}. \\ \hline 
Chim et al. (2015) \cite{134} & - Smart grid network based on hierarchical architecture, i.e., HANs, BANs, NANs. & - Providing the privacy-preserving recording and gateway-assisted authentication. & - Preparation phase;\newline - Power plan submission phase;\newline - Power plan processing phase;\newline - Reconciliation phase;\newline - System master secret updating phase. & + The message filtering at gateway smart meters can helpful in reducing the impact of attacking traffic.\newline + The privacy-preserving  and traceability is considered.\newline - No comparison with other schemes.\newline - Distributed denial of service (DDoS) attacks is not considered. \\ \hline 
Mahmood et al. (2016) \cite{135} & - The system model is homogeneous to the model in ref. \cite{128}. & - Detect and omit some attacks, namely, replay, false message injection, message analysis and modification attacks. & - Initialization;\newline - Authentication;\newline - Message transmission. & + Efficient in terms of communication cost and computation cost compared to the schemes \cite{140} \cite{166}.\newline + Resistance to attacks, namely, replay, false message injection, message analysis and modification attacks.\newline + The reports' confidentiality and integrity are considered.\newline - Location privacy is not considered.\newline \newline  \\ \hline 
\end{supertabular}}
\label{tab:Tab5e}

\subsection{Authentication protocols for M2M}

The surveyed papers of authentication protocols for Machine to machine communications (M2M) as shown in Tab. 20 are published between 2012 and 2016. In order to speed up the process of authentication and  avoid authentication signaling overload, Lai et al. \cite{76} focused on the problem of group authentication and key agreement for resource-constrained M2M devices in 3GPP networks.  Specifically, the authors proposed a novel group-based lightweight authentication scheme for resource constrained M2M, called GLARM. The network model used in \cite{76} is based on 3GPP standard with three domains, including, access networks, evolved packet core, and non-3GPP domain, e.g., Internet. To guarantee the entity mutual authentication and secure key agreement, the GLARM scheme uses two main phases, namely, 1) Initialization phase and 2) Group authentication and key agreement phase. In addition, the GLARM scheme can not only ensure QoS for machine-type communications devices, but the computation complexity is much less than schemes \cite{77},\cite{78} and \cite{79}. In order to distinguish between different physical devices running the same software and detecting mimic attacks, Chen et al. \cite{71} proposed an authentication protocol for the IoT, named S2M. The S2M protocol uses tree main phases, namely, 1) audio-handshake phase, 2) mixed-signal generation phase, and 3) feature extraction and storage phase. S2M can achieve variable distance authentication and active attack detection using acoustic hardware (Speaker/Microphone) fingerprints. In addition, S2M is efficient in terms of lower error rates compared with DISWN \cite{72}, LDTLS \cite{73}, PLTEA \cite{74}, and SeArray \cite{75}, but the performance of the methods in terms of privacy preservation is not analyzed, especially in comparison to the GLARM scheme \cite{76}.

To authenticate a group of devices at the same time, Lai et al. \cite{78} proposed a scheme named, SEGR. Based on roaming phase, SEGR can achieving mutual authentication and key agreement between all Machine-type Communication (MTC) devices when a group of MTC devices roams between 3GPP and WiMAX networks. SEGR is efficient in terms of the communication overhead computation complexity compared to the scheme in \cite{81} and the scheme without aggregation, but again a comparison with other methods such as the GLARM scheme \cite{76} regarding  privacy preservation is missing. We also note that resistance to attacks of the SEGR method is not studied in the  article as well \cite{78}. To guarantee privacy-preservation and key forward/backward secrecy, Lai et al. \cite{79} proposed an efficient group authentication and key agreement protocol, called SE-AKA, which is based on authentication and key agreement (AKA) protocol. The overhead of authentication message delivery of SE-AKA is lower than other existing AKA protocols, but the computational overhead is larger than that of other traditional protocols such as the work \cite{84}. In addition, SE-AKA has smaller storage costs than others AKA protocols. Similar to the SE-AKA protocol, Lai et al. in \cite{136} proposed a lightweight group authentication protocol for M2M, called LGTH, which is efficient in terms of the signaling and computation overhead compared to the schemes \cite{77} and \cite{146}. Similar to the SE-AKA \& LGTH protocols, Fu et al. \cite{81} proposed a group-based handover authentication scheme for mobile WiMAX networks. Based on the handover authentication phase, the work \cite{81} is efficient in terms of the computational and communication overhead compared to three schemes \cite{87},\cite{88}, and \cite{84}, but the resistance to attacks is not studied and no threat model is presented. 

In order to achieve a mutual authentication process in machine-to-machine home network service, Sun et al. \cite{117} proposed an M2M application model for remote access to the intelligence home network service using the existing Time Division-Synchronous Code Division Multiple Access (TD-SCDMA) system. The protocol \cite{117} is efficient in terms of the amount of calculations needed and communication volume compared to the protocol in \cite{119}, but the article lacks a comparison of performance in terms of non-repudiation against other schemes such as the PBA \cite{92}. To achieve the authentication of mobile subscribers in the roaming service, Lai et al. \cite{118} proposed a conditional privacy-preserving authentication with access linkability, called CPAL. The CPAL can 1) provide a strong anonymous access authentication, 2) guarantee user tracking on a disputed access request, and 3) achieve anonymous user linking and efficient user revocation for dynamic membership. The CPAL is efficient in terms of communication overhead and computation cost compared to two strong anonymous schemes \cite{121} and \cite{122}, but privacy aspects are not analyzed such as location privacy. Without adding any extra hardware devices, Zhu et al. \cite{124} proposed a dual-factor authentication scheme, called Duth, designed for Android smartphone devices. Based on two main processes, namely, 1) feature-set extraction and storing for registration, 2) dual-factor authentication, the Duth scheme can satisfy the user-friendly requirements, along with a reasonable false rejection rate, providing on the same time an authentication process for Android smartphone devices.
\\
\begin{center}
\topcaption{Summary of authentication protocols for (IoV) (Published between 2013 and 2016)}
\end{center}
{\tiny
\begin{supertabular} {|p{0.5in}|p{1in}|p{1in}|p{1in}|p{3in}|} \hline  
\textbf{Protocol } & \textbf{Network model} & \textbf{Goals} & \textbf{Main processes} & \textbf{Performances (+) and limitations (-)} \\ \hline 
Céspedes et al. (2013) \cite{89} & - A vehicular communications network with Access Routers (ARs) that connect the VANET to external IP networks & - Achieving mutual authentication against authentication attacks & - Key establishment phase;\newline - MR registration phase;\newline - Authentication phase;\newline - Mobile router revocation. & + Considers the asymmetric links in the VANET.\newline + Achieving less location update cost compared with the scheme \cite{96}.\newline + The handover delay lower than the one in the scheme \cite{96}.\newline + Resistance to replay attack, man-in-the-middle attack, and denial of service (DoS) attack.\newline - Privacy-preserving is not analyzed compared to the GLARM scheme \cite{76}.\newline - Lack non-repudiation compared to the PBA scheme in \cite{92}. \\ \hline 
Wasef et al. (2013) \cite{90}\newline  & - VANET with a trusted authority and some entities including On-Board Units (OBUs), and infrastructure Road-Side Units (RSUs). \newline - Vehicle-to-Vehicle (V2V) and Vehicle-to-Infrastructure (V2I) communications & - Expedite message authentication & - System initialization;\newline - Message authentication;\newline - Revocation. & + Efficient in terms of computational complexity of revocation status checking.\newline + The authentication delay is constant and independent of the number of revoked certificates.\newline + Efficient in terms of communication overhead.\newline + Resistance to forging attack, replay attack, and colluding attack.\newline - Storage costs is not considered.\newline - No comparison with other schemes.\newline - Lack non-repudiation compared to the PBA scheme in \cite{92}. \\ \hline 
Shao et al. (2016) \cite{91}\newline  & - VANET with some parties, including, central authority, tracing manager, many RSUs, and many OBUs. & - Guarantee unforgeability, anonymity, and traceability. & Initialization stage;\newline Registration stage;\newline Join stage;\newline Sign stage;\newline Verify stage;\newline Trace stage. & + Efficient in terms of the computational cost of three operations, namely, Initialization, Registration, and Trace.\newline + Can prevent replay attacks.\newline - No comparison with other schemes.\newline - The communication overhead is not studied.\newline - Lack non-repudiation compared to the PBA scheme in \cite{92}. \\ \hline 
Lyu et al. (2016) \cite{92}\newline  & - VANET with divide messages into two types 1) single-hop beacons and 2) multi-hop traffic data. & - Guarantee some properties such as timely authentication, non-repudiation, packet losses resistant, and DoS attacks resistant. & - Chained keys generation;\newline - Position prediction;\newline - Merkle hash tree construction;\newline - Signature generation;\newline - Self-Generated MAC Storage;\newline - Signature Verification. & + Considers the non-repudiation.\newline + The computational cost reduces with the increasing of time frame.\newline + Can resist packet losses.\newline + Maintain high packet processing rate with low storage overhead.\newline + Efficient in terms of overall delay compared to the TESLA scheme in \cite{103} \cite{104} \cite{105} and the VAST scheme in \cite{106}.\newline - Privacy-preserving is not analyzed compared to the GLARM scheme \cite{76}. \\ \hline 
Zhang et al. (2016) \cite{93}\newline  & - Trusted authority (TA), a number of RSUs and vehicles & - Guarantee the conditional unlinkability, ideal tamper-proof device (TPD) freeness, key escrow freeness. & - Member secrets generation;\newline - Vehicle sign;\newline - Message verification and signature storage;\newline - Trace internal pseudo-identity (IPID) and authentication key update;\newline - On-Line update. & + Efficient in terms of message authentication delay on average.\newline + Considers privacy preserving.\newline + Resistance to the side-channel attack, false messages attack, denial-of-service (DoS) attack, and Sybil attack.\newline + Efficient than the ECDSA protocol in \cite{107} and more efficient than the IBA scheme in \cite{108} on average.\newline - Lack non-repudiation compared to the PBA scheme in \cite{92}. \\ \hline 
Zhang et al. (2015) \cite{108} & - VANET with four main entities, i.e., key generator center (KGC), traffic management authority (TMA), RSUs and vehicles. & - Guarantee some properties such as message authentication, non-repudiation, message confidentiality, privacy, and traceability. & - System setup;\newline - Protocol for STP and STK distribution;\newline - Protocol for common string synchronization;\newline - Protocol for vehicular communications. & + Efficient in terms of the average message delay and the verification delay.\newline + Efficient in terms of  verification delay compared to the scheme in \cite{110}.\newline + Considers the non-repudiation.\newline + Resistance to attacks, namely, message reply, message modification, movement tracking.\newline - Location privacy is not considered. \\ \hline 
Dolev et al. (2016) \cite{94}\newline  & - The vehicle network is divided into the controller area network (CAN), local interconnect network (LIN), and media oriented system (MOST). & - Ensure the countermeasures against the Man-in-the-Middle attack under the vehicle authentication. & - System settings;\newline - Certificate authority;\newline - Vehicular attributes. & + Efficient in terms of  iteration cost compared to the existing Authenticated Key Exchange (AKE) protocols such as ISO-KE \cite{112} and SIGMA \cite{113}.\newline + Resistance to attacks, namely, Man-in-the-Middle attack and impersonation attack.\newline - Privacy-preserving is not analyzed compared to the GLARM scheme \cite{76}. \\ \hline 
Chan et al. (2014) \cite{95} & - Smart grid electric vehicle ecosystem & - Provides assurance of the digital identity and the device's controllability in the physical domain. & - Communication settings;\newline - Cyber-physical device authentication. & + Resistance to substitution attacks.\newline - No comparison with other schemes.\newline - The average message delay and the verification delay are not evaluated. \\ \hline 
Lai et al. (2015) \cite{125} & - The trust authority (TA), road side units (RSUs), and vehicles. & - Provide strong anonymous access authentication. & - System initialization;\newline - Platoon merging;\newline - Platoon splitting;\newline - Anonymous authentication with traceability. & + Efficient in terms of  computational cost.\newline - No comparison with other schemes.\newline - The average message delay and the verification delay are not evaluated.\newline - No threat model presented. \\ \hline 
\end{supertabular}}
\label{tab:Tab5d}

\begin{center}
\topcaption{Summary of authentication protocols for IoS (Published in 2016)}
\end{center}
{\tiny
\begin{supertabular} {|p{0.5in}|p{1in}|p{1in}|p{1in}|p{3in}|} \hline 
\textbf{Protocol } & \textbf{Network model} & \textbf{Goals} & \textbf{Main processes} & \textbf{Performances (+) and limitations (-)} \\ \hline 
Kumari et al. (2016) \cite{167} & - Wireless sensor network (WSN) with the service seeker users, sensing component sensor nodes (SNs) and the service provider base-station or gateway node (GWN). & - Providing mutual authentication with forward secrecy and wrong identifier detection mechanism at the time of login. & - Initialization phase;\newline - User registration phase;\newline - Login phase;\newline - Authentication \& key agreement phase;\newline - Password change phase; & + The user is anonymous.\newline + Resistance to attacks, namely, user impersonation attack, password guessing attack, replay attack, stolen verifier attack, smart card loss attack, session-specific temporary information attack, GWN Bypass attack, and privileged insider attack.\newline + Provides a secure session-key agreement and forward secrecy.\newline + Provides freely password changing facility.\newline + Efficient in unauthorized login detection with wrong identity and password.\newline - The data integrity is not considered. \\ \hline 
Chung et al. (2016) \cite{168} & - Wireless sensor networks for roaming service.  & - Providing an enhanced lightweight anonymous authentication to resolve the security weaknesses of the scheme \cite{197}. & - Registration phase;\newline - Login and authentication phase;\newline - Password change phase. & + Considers anonymity, hop-by-hop authentication, and untraceability.\newline + Resistance to attacks, namely, password guessing attack, impersonation attack, forgery attack, known session key attack, and fair key agreement.\newline - Location privacy is not considered. \\ \hline 
Jan et al. (2016) \cite{169} & - A cluster-based hierarchical WSN. & - Providing an extremely lightweight payload-based mutual authentication & - Token-based cluster head election;\newline - Payload-based mutual authentication. & + Considers authenticity, confidentiality, and freshness of data for node-to-node communication.\newline + Robustness against attacks, namely, replay attack, resource exhaustion attack, and Sybil attack.\newline + Efficient in terms of average energy consumption and Handshake duration compared to the LEACH-C scheme in \cite{198} and the SecLEACH scheme \cite{199}.\newline - Privacy-preserving is not analyzed compared to the GLARM scheme \cite{76}.\newline - Lack non-repudiation compared to the PBA scheme in \cite{92}.\newline - Storage costs is not considered. \\ \hline 
Amin et al. (2016) \cite{170} & - Three types of entities such as (a) sensor nodes, (b) gateway nodes, and (c) users. & -  Providing a user authentication and key agreement protocol to resolve the security weaknesses of the scheme \cite{202}. & - System setup phase;\newline - Sensor node registration phase;\newline - User registration phase;\newline - Login phase;\newline - Authentication and key agreement phase;\newline - Dynamic node addition phase;\newline - Password update phase. & + Resistance to attacks, namely, password guessing attack, user impersonation attack, gateway node impersonation attack, privileged insider attack, replay attack, and session key computation attack.\newline + Efficient in terms of computational cost compared to the schemes \cite{202} \cite{203} \cite{204} \cite{205} \cite{206}.\newline + Efficient in term of storage and communication cost compared to the schemes \cite{202} \cite{203} \cite{204} \cite{205} \cite{206}.\newline - Lack non-repudiation compared to the PBA scheme in \cite{92}. \\ \hline 
Gope et al. (2016) \cite{171} & - Real-time data access in WSNs. & - Ensuring the user anonymity, perfect forward secrecy, and resiliency of stolen smart card attacks. & - Registration phase;\newline - Anonymous authentication and key exchange phase;\newline - Password renewal phase;\newline - Dynamic node addition phase. & + Considers the user anonymity and untraceability.\newline + Provides perfect forward secrecy.\newline + Security assurance in case of lost smart card.\newline + Resilience against node capture attack and key compromise impersonation Attack.\newline + Efficient in terms of computational and communication cost compared to the schemes \cite{206}, \cite{203}, \cite{207}, \cite{208} and \cite{172}.\newline - The average message delay and the verification delay are not evaluated. \\ \hline 
Das (2016) \cite{172} & - WSN  & - Providing a user authentication to resolve the security weaknesses of the scheme \cite{208}. & - Pre-deployment of sensor nodes phase;\newline - Registration phase;\newline - Login phase;\newline - Authentication and key agreement phase;\newline - Password and biometric update phase;\newline - Dynamic node addition phase. & + Resistance to attacks, namely, privileged insider attack, offline/online password and biometric key guessing attack, replay attack, man-in-the-middle attack, stolen-verifier attack, forgery attack, and node capture attack.\newline + Efficient in terms of computational and communication overhead compared to the schemes \cite{203} \cite{209} \cite{210}.\newline - Lack non-repudiation compared to the PBA scheme in \cite{92}. \\ \hline 
Chang et al. (2016) \cite{173} & - Users, sensor nodes, and gateway node in WSN.  & - Providing mutual authentication and perfect forward secrecy. & - Registration phase;\newline - Authentication phase;\newline - Password changing phase. & + Considers the session key security, perfect forward secrecy, and user anonymity.\newline + Resistance to attacks, namely, replay attack and smart card lost attack.\newline + Efficient in terms of computation cost in the authentication phases compared to the schemes \cite{202} \cite{195} \cite{212} \cite{213}.\newline - Privacy-preserving is not analyzed compared to the GLARM scheme \cite{76}. \\ \hline 
Jiang et al. (2016) \cite{174} & - Users, sensor nodes, and gateway node in WSN.  & - Providing mutual authentication, anonymity, and untraceability. & - Registration phase;\newline - Login and authentication phase. & + Provides mutual authentication, session key agreement, user anonymity, and user untraceability.\newline + Resistance to attacks, namely, smart card attack, impersonation attack, modification attack, man-in-the-middle attack, and tracking attack.\newline + Efficient in terms of computational cost compared to the schemes \cite{196} \cite{206} \cite{212} \cite{213} \cite{214}.\newline - Wormhole attack and blackhole attack are not considered. \\ \hline 
Farash et al. (2016) \cite{175}\newline  & - Users, sensor nodes, and gateway node in WSN.  & - Providing the user authentication with traceability protection and sensor node anonymity. & - Pre-deployment phase;\newline - Registration phase;\newline - Login and authentication phase;\newline - Password change phase. & + Efficient in terms of communication, computation and storage cost compared to the scheme \cite{202}.\newline + Resistance to attacks, namely, replay attack, privileged-insider attack, man-in-the-middle attack, insider and stolen verier attack, smart card attack, impersonation attack, bypassing attack, many logged-in users with the same login-id attack, password change attack, and DoS attack.\newline - Wormhole attack and blackhole attack are not considered. \\ \hline 
Srinivas et al. (2017) \cite{176}\newline  & - Users, sensor nodes, and gateway node in WSN.  & - Providing the mutual authentication with anonymity and unlinkability. & - System setup phase;\newline - Sensor node registration phase;\newline - User registration phase;\newline - System environment phase;\newline - Login, authentication, and key agreement phase;\newline - Dynamic node addition phase;\newline - Password change phase. & + Efficient in terms of communication overhead during the login and authentication phase compared to the schemes \cite{170} \cite{215} \cite{216} \cite{206} \cite{204} \cite{205} \cite{203} \cite{218} \cite{202}.\newline + Resistance to attacks, namely, replay attack, privileged insider attack, session key computation attack, gateway node impersonate attack, user impersonate attack, guessing attack, and sensor node spoofing attack.\newline - Wormhole attack and blackhole attack are not considered.\newline - Privacy-preserving is not analyzed compared to the GLARM scheme \cite{76}. \\ \hline 
Kumari et al. (2016) \cite{177} & - Users, sensor nodes, and gateway node in WSN.  & - Providing the mutual authentication with traceability and anonymity. & - Offline sensor node registration phase;\newline - User registration phase;\newline - Login phase;\newline - Authentication and key agreement phase;\newline - Password update phase;\newline - Dynamic sensor node addition phase. & + Efficient in terms of end-to-end delay (EED) (in seconds) and throughput (in bps).\newline + Efficient in terms of computation cost in login and authentication phases compared to both schemes Turkanovic et al. \cite{202} and Farash et al. \cite{175}.\newline + Resistance to attacks, namely, replay attack, stolen smart card attack, privileged-insider attack, offline password guessing attack, impersonation attack, and sensor node capture attack.\newline - Wormhole attack and blackhole attack are not considered.\newline - Lack non-repudiation compared to the PBA scheme in \cite{92}.\newline  \\ \hline 
Sun et al. (2016) \cite{178} & - Multicast communications in WSNs, including, sink and many groups, and each group have a powerful node and many low ordinary nodes.  & - Providing the broadcast authentication and enhanced collusion resistance & - Initialization;\newline - Broadcast;\newline - Group keys' recovery and pairwise keys' updating;\newline - Node addition;\newline - Node revocation. & + Collusion resistance.\newline + Resistance to attacks, namely, PKE-attack and PF-attack.\newline + Efficient in terms of storage, computation, and communication overhead compared to the schemes \cite{220} \cite{221} \cite{222} \cite{223}.\newline - The end-to-end delay and throughput are not evaluated compared to the scheme \cite{177}.\newline - Replay attack is not considered. \\ \hline 
Jiang et al. (2016) \cite{179}\newline  & - Users, sensor nodes, and gateway node in WSN.  & - Achieving mutual authentication among the communicating agents with user anonymity and untraceability. & - Registration phase;\newline - Login phase;\newline - Authentication phase;\newline - Password change phase. & + Resistance to attacks, stolen-verifier attack, guessing attack, impersonation attack, modification attack, man-in-the-middle attack, and replay attack.\newline + Efficient in terms of computational cost compared to the schemes in \cite{206} \cite{212} \cite{213} \cite{214}.\newline - The end-to-end delay and throughput are not evaluated compared to the scheme \cite{177}.\newline - Collusion resistance is not considered compared to the scheme \cite{178}.  \\ \hline 
\end{supertabular}}
\label{tab:Tab5f}

\subsection{Authentication protocols for IoV}

The surveyed papers of authentication protocols for Internet of Vehicles (IoV) as shown in Tab. 21 are published between 2013 and 2016. Céspedes et al. in \cite{89} considered the security association between asymmetric links during Vehicle to Vehicle (V2V) communications. More precisely, the authors proposed a multi-hop authenticated proxy mobile IP scheme, called MA-PMIP. Based on authentication phase and mobile router revocation, MA-PMIP can achieve less location update cost compared with the scheme \cite{96} and the handover delay lower than the scheme \cite{96}. In addition, MA-PMIP can achieve mutual authentication against authentication attacks but the privacy preserving is not analyzed compared to the GLARM scheme \cite{76}. In order to expedite message authentication in VANET, Wasef et al. \cite{90} proposed an expedite message authentication protocol, named EMAP. Based on the revocation checking process, EMAP can overcome the problem of the long delay incurred in checking the revocation status of a certificate using a certificate revocation list. EMAP is efficient in terms of computational complexity of revocation status checking and the authentication delay is constant and independent of the number of revoked certificates. Therefore, the question we ask here is: can these protocols work well in the decentralized group model? The authentication scheme proposed recently by Shao et al. in \cite{91} can answer this question where he can achieve two requirements for threshold authentication, namely, distinguishability and efficient traceability. The protocol in \cite{91} is proven that is secured by three theorems, namely, 1) The proposed group signature scheme satisfies unforgeability, 2) The proposed group signature scheme satisfies anonymity, and 3) The proposed theorem satisfies the traceability.

To achieve the non-repudiation in IoV, Lyu et al. in \cite{92} proposed a lightweight authentication scheme called PBA. Based on the idea of merkle hash tree construction and self-generated MAC storage, the PBA scheme can resist packet losses and maintain high packet processing rate with low storage overhead. The PBA is efficient in terms of overall delay compared to the TESLA scheme in \cite{103}, \cite{104}, and \cite{105} and the VAST scheme in \cite{106}. Zhang et al. in \cite{108} considers a VANET with four main entities, i.e., key generator center (KGC), traffic management authority (TMA), RSUs and vehicles.  Based on identity-based aggregate signatures, the protocol in \cite{108} can guarantee some properties such as message authentication, non-repudiation, message confidentiality, privacy, and traceability. Similar to the scheme \cite{108}, Zhang et al. \cite{93} proposed an efficient distributed aggregate privacy preserving authentication protocol, called DAPPA, which is based on a new security tool called multiple-TA OTIBAS (MTA-OTIBAS). The DAPPA protocol can guarantee the conditional unlinkability, ideal tamper-proof device (TPD) freeness, key escrow freeness. In addition, the DAPPA protocol is efficient than the ECDSA protocol in \cite{107} and more efficient than the IBA scheme in \cite{108} on average, but lack non-repudiation compared to the PBA scheme in \cite{92}. Based on monolithically certified public key and attributes, Dolev et al. \cite{94} proposed an idea to ensuring the countermeasures against the Man-in-the-Middle attack under the vehicle authentication. The work in \cite{94} is efficient in terms of  iteration cost compared to other existing Authenticated Key Exchange (AKE) protocols such as ISO-KE \cite{112} and SIGMA \cite{113}. To defend against coordinated cyber-physical attacks, Chan et al. \cite{95} proposed a two-factor cyber-physical device authentication protocol, which can be applied in the IoV. Especially in the IoT,  the vehicles may join or leave the platoon at any time in the platoon-based vehicular cyber-physical system. To guarantee anonymity of platoon members, Lai et al. \cite{125} proposed a secure group setup and anonymous authentication scheme, named SGSA, for platoon-based vehicular cyber-physical systems. Based on the anonymous authentication with traceability phase, the SGSA scheme can provide strong anonymous access authentication.

\subsection{Authentication protocols for IoE}

The surveyed papers of authentication protocols for Internet of Energy (IoE) as shown in Tab. 22 are published between 2011 and 2016. We noted here that we have reviewed some authentication protocols proposed for secure smart grid communications in our survey in \cite{149}, namely, the schemes in \cite{137}, \cite{138}, \cite{139}, \cite{140}, \cite{141}, \cite{142}, \cite{143}, \cite{144}, \cite{145}. In this subsection, we will review only the works that are not reviewed in the survey \cite{149}.

To provide multicast authentication in smart grid, Li et al. \cite{127} proposed the scheme Tunable Signing and Verification (TSV). Specifically, TSV combines Heavy Signing Light Verification (HSLV) and Light Signing Heavy Verification (LSHV) to achieve a flexible tradeoff between the two. TSV can reduce the storage cost, but the privacy-preserving is not discussed and the reports' confidentiality and integrity are not considered compared to the scheme \cite{128}. The smart meters is planning to reduce the time intervals to 1 min or even less. For this, Li et al. \cite{128} developed a merkle-tree-based authentication scheme to minimize computation overhead on the smart meters. The work \cite{128} is efficient in terms of  computation complexity of the HAN user and the neighborhood gateway compared to the Rivest--Shamir--Adleman (RSA)-based authentication scheme \cite{151}. Therefore, Li et al. \cite{129} fixed the single-point failure in smart grid by proposing the idea of deploys a fault tolerance architecture to execute the authentication approach without any additional configuration or setup. Based on both main processes, namely, 1) Batch verification and trinary diagnose TreeBatch, and 2) Signature amortization for Package Blocks, the work \cite{129} can legalize the data aggregation with tremendously less signing and verification operations.

Nicanfar et al. \cite{130} addressed the key management for unicast and multicast communications in the smart grid. The work \cite{120} proposed a scheme for the mutual authentication between the smart grid utility network and Home Area Network smart meters, called SGAS-I, which he can increases performance of the key management and does not cause any security drawback. Based on the multicast key support phase, SGAS-I can provide simplicity and low overhead, but the reports' confidentiality and integrity are considered compared to the scheme \cite{128}. To guarantee the message authentication with identity privacy and traceability, Chim et al. \cite{131} proposed a scheme, called PASS, for the hierarchical structure of a smart grid. The PASS scheme focus only on the substation-to-consumer subsystem where the real identity of any smart appliance can only be known by the control center using the concept of pseudo identity. Similar to the PASS scheme, Fouda et al. \cite{132} proposed a scheme that can only providing an authenticated and encrypted channel for the late successive transmission but can also establish a semantic-secure shared key in the mutual authentication environment. The work in \cite{132} is efficient in terms of communication overhead and message decryption/verification delay compared to ECDSA-256, but the identity privacy and traceability are not considered compared to the scheme \cite{131}.

In order to provide the mutual authentication between smart meters and the security and authentication server in the smart grid using passwords, Nicanfar et al. \cite{133} proposed a mutual authentication scheme and a key management protocol, called SGMA and SGKM, respectively. The SGMA scheme concentrates on data communications over the advanced metering infrastructure (AMI) outside of the HAN domain, which each node has a unique ID and each smart meter has a unique serial number SN embedded by the manufacturer and an initial secret password. On the other hand, the SGKM protocol concentrates on node-to-node secure communications, which the nodes have the appropriate private--public keys to be used for unicast. Based on the multicast key mechanism, the SGMA scheme can prevent various attacks while reducing the management overhead, but lack non-repudiation compared to the PBA scheme in \cite{92}. Shim et al. \cite{134} consider a smart grid network based on hierarchical architecture, i.e., HANs, BANs, NANs. The work \cite{134} proposed privacy-preserving recording and gateway-assisted authentication of power usage information. The message filtering at gateway smart meters can helpful in reducing the impact of attacking traffic. Similar to the scheme \cite{134}, Mahmood et al. \cite{135} proposed a lightweight message authentication scheme. Based on two main processes, namely, 1) Authentication and 2) Message transmission, the scheme \cite{135} can detect and omit some attacks, namely, replay, false message injection, message analysis and modification attacks. In addition, the scheme \cite{135} is efficient in terms of communication cost and computation cost compared to the schemes \cite{140} and \cite{166}, but the location privacy is not considered.

\subsection{Authentication protocols for IoS}

The surveyed papers of authentication protocols for Internet of Sensors (IoS) as shown in Tab. 23 are published in 2016. We noted here that we have reviewed some authentication protocols proposed for ad hoc social network (an application of WSN) in our survey in \cite{233}, namely, the protocols in \cite{234}, \cite{235}, \cite{236}, \cite{237}, \cite{238}, and \cite{239}. In this subsection, we will review only the works that are not reviewed in the survey \cite{233} and the articles published in 2016 related to authentication protocols for IoS. For more details about the articles published before 2016, we refer the reader to six surveys published in 2013, 2014, and 2015, namely, \cite{186}, \cite{187}, \cite{188}, \cite{189}, \cite{190}, \cite{191}.

Kumari et al. \cite{167} reviewed and examined both schemes proposed by Li et al.'s in \cite{195} and He et al.'s in \cite{196} for its suitability to WSNs. Based on the results of this analysis, the authors proposed a chaotic maps based user friendly authentication scheme for WSN with forward secrecy and wrong identifier detection mechanism at the time of login. The idea is to establish a session key between user and sensor node (SN) using extended chaotic maps. The scheme of Kumari et al. \cite{167} is efficient in unauthorized login detection with wrong identity and password, but the data integrity is not considered.  Similar to the \cite{167}, Chung et al. \cite{168} reviewed and examined the scheme \cite{197}. Based the security weaknesses of the scheme \cite{197}, the work \cite{168} proposed an enhanced lightweight anonymous authentication scheme for a scalable localization roaming service in WSN. Using three phases, namely, 1) Registration phase, 2) Login and authentication phase, and 3) Password change phase, the work \cite{168} can provide both anonymity, hop-by-hop authentication, and untraceability, but location privacy is not considered.

Jan et al. \cite{169} proposed an extremely lightweight payload-based mutual authentication, called PAWN, for the cluster-based hierarchical WSN. The PAWN scheme is based on two main phases, namely, 1) Token-based cluster head election and 2) Payload-based mutual authentication. With the phase 1, the higher-energy nodes perform various administrative tasks such as route discovery, route maintenance, and neighborhood discovery. The authentication procedure is accomplished using the cooperative neighbor $\times$ neighbor (CNN) \cite{201}, i.e., session initiation, server challenge, client response and challenge, server response. The PAWN scheme is efficient in terms of average energy consumption and Handshake duration compared to the LEACH-C scheme in \cite{198} and the SecLEACH scheme \cite{199}, but the privacy-preservation is not analyzed compared against other methods, such as the GLARM scheme \cite{76}. Based on the security weaknesses of the scheme \cite{202}, Amin et al. \cite{170} proposed a secure lightweight scheme for user authentication and key agreement in multi-gateway based WSN. The scheme \cite{170} is efficient in terms of computational cost, storage and communication cost compared to the schemes \cite{202}, \cite{203} \cite{204}, \cite{205}, and \cite{206}. In addition, the scheme \cite{170} can provide very less energy consumption of the sensor nodes and user anonymity. 

For the security of real-time data access in WSNs, Gope et al. \cite{171} proposed an authentication protocol to ensuring the user anonymity, perfect forward secrecy, and resiliency of stolen smart card attacks. The protocol \cite{171} is efficient in terms of computational and communication cost compared to the schemes \cite{206}, \cite{203}, \cite{207}, \cite{208} and \cite{172}. Based on the security weaknesses of the scheme \cite{208}, Das \cite{172} proposed a secure and robust temporal credential-based three-factor user authentication scheme. The scheme \cite{172} use a biometric, password and smart card of a legal user. The simulation results of the scheme \cite{172} demonstrate that it is efficient in terms of computational and communication overhead compared to the schemes \cite{203}, \cite{209}, and \cite{210}. Based on the weaknesses in Turkanovic et al.'s protocol \cite{202}, Chang et al. \cite{173} proposed a flexible authentication protocol using the smart card for WSNs, which operates in two modes, namely, 1) provides a lightweight authentication scheme, and 2) an advanced protocol based on ECC, which provides perfect forward secrecy. Both this two modes are efficient in terms of computation cost in the authentication phases compared to the schemes \cite{202}, \cite{195}, \cite{212}, and \cite{213}. 

Trying to deal with the weaknesses of the scheme presented in \cite{196}, Jiang et al. \cite{174} proposed an untraceable two-factor authentication scheme based on elliptic curve cryptography. The scheme \cite{174} is efficient in terms of computational cost compared to previous schemes \cite{196}, \cite{206}, \cite{212}, \cite{213} , and \cite{214}, but the performance of the system under common attacks such as the wormhole attack and the blackhole attack is not presented. Based on the weaknesses in the scheme \cite{202}, Farash et al. \cite{175} proposed an efficient user authentication and key agreement scheme for heterogeneous wireless sensor network tailored for the Internet of Things environment. The scheme \cite{175} is efficient in terms of communication, computation and storage cost compared to the scheme \cite{202}, but again the performance of the system under the wormhole attack or the blackhole attack is not presented. Based on the weaknesses in the Amin and Biswas's scheme \cite{170}, Srinivas et al. \cite{176} proposed a user authentication scheme for multi-gateway WSNs. The scheme \cite{176} is efficient in terms of communication overhead during the login and authentication phase compared to the schemes \cite{170}, \cite{215}, \cite{216} \cite{206}, \cite{204}, \cite{205}, \cite{203}, \cite{218}, \cite{202}, but the performance of the system in terms of  privacy-preservation is not analyzed compared to previous methods, such as the GLARM scheme \cite{76}. Similar to both schemes \cite{174} and \cite{176}, Kumari et al. \cite{177} pointed out that the scheme of Farash et al. \cite{175} is insecure against some attacks. Especially, the work presented in \cite{177} is efficient not only in terms of end-to-end delay (EED) (in seconds) and throughput (in bps), but also in terms of computation cost in either login and authentication phases compared to both schemes Turkanovic et al. \cite{202} and Farash et al. \cite{175}. 

Sun et al. \cite{178} considered the multicast communications in WSNs, including, sink and many groups, where each group may have a powerful node and many low ordinary nodes. The powerful node acts  as the group manager (GM), and is responsible for network security management, such as key issues, updating, revocation, intrusion detection, etc. Then, the authors reviewed and examined the scheme \cite{224} in order to propose a scheme that considers the forward security, backward security, and collusion resistance. Based on the idea of access polynomial, the Sun et al. scheme \cite{178} is efficient in terms of storage, computation, and communication overhead compared to the schemes \cite{220}, \cite{221}, \cite{222} , and \cite{223}, but not only the end-to-end delay and throughput are not evaluated compared to the scheme \cite{177}, also the replay attack is not considered. Jiang et al. proposed a scheme \cite{179} that can achieve mutual authentication among the communicating agents with user anonymity and untraceability. In addition, the Jiang et al. scheme \cite{179} is efficient in terms of computational cost compared to the schemes in \cite{206}, \cite{212}, \cite{213}, \cite{214}, but the collusion resistance is not considered compared to the scheme in \cite{178}. 

Based on the weaknesses in the scheme \cite{227}, Wu et al. \cite{180} proposed an improved three-factor authentication scheme for WSNs, which can be resistant to the de-synchronization attack. Das et al. \cite{181} reviewed the recently proposed Chang--Le's two protocols \cite{228} and then showed that their protocols are insecure against some known attacks. Liu and Chung \cite{182} proposed a secure user authentication scheme for wireless healthcare sensor networks, which is efficient in terms of computation cost of compared to both schemes in \cite{229} and \cite{230}. Gope et al. \cite{183} proposed a special idea for resilience of DoS attacks in designing anonymous user authentication protocol. Combining three techniques, namely, smart card, password, and personal biometrics, Das et al. \cite{184} proposed a three-factor user authentication and key agreement scheme based on multi-gateway WSN architecture. The scheme \cite{184} is efficient in terms of computational, communication, and energy costs compared to the schemes \cite{170}, \cite{205}, \cite{203}, \cite{202}. Benzaid et al. \cite{185} proposed an accelerated verification of digital signatures generated by BNN-IBS \cite{231}, which is an idea inspired by the acceleration technique of Fan and Gong \cite{232}.

\section{Open Issues}\label{sec:open-issues}
\subsection{M2M Open Issues}
M2M communications can facilitate many applications, like e-health, smart grids, industrial automation and enviromental monitoring but on the same time face various security threats and trust issues. Especially in e-health authentication of the devices must be robust to attacks that could threaten the correct exhancge of information and the consequently the life of the patient. In order to safely share and manage access to information in the healthcare system, it is essential to be able to authenticate users, including organizations and people. In Australia authentication is achieved through the use of digital certificates that conform to the Australian Government endorsed Public Key Infrastructure (PKI) standard, through the  National Authentication Service for Health (NASH) but thorough research of the resistance to attacks of this and other similar systems is needed in order to reassure its robustness. Scalability and Heterogeneity is a rather general problem when dealing with M2M communication of devices that come from different vendors and using different operating systems. Solutions that focus only to Android devices \cite{124}  cannot guarantee end to end security of the system. 

\subsection{IoV Open Issues}

Although a number of authentication protocols have been proposed recently which are capable of guaranteeing authentication for a network of vehicles, there are still open issues that need to be addressed by the research community.

\subsubsection{Autonomous driving} Until now anonymity of platoon members has been addressed in \cite{125}, which is capable of providing strong anonymous access authentication to the members of the platoon. Taking on step further and dealing with full automated vehicles that will be able to create platoons on the fly, with no central entity or trust authority in reach, novel authentication methods that vehicles can run by themselves must be developed. This could be done using several techniques. One method would be to use digital signatures, where each vehicle holds its own signing key and can verify its identity by signing challenges, combined with a defense mechanism that can face MITM attacks. Other method could be the use of the trust levels of every vehicle using methods similar to \cite{276}. 

\subsubsection{Heterogeneous vehicular networking} The design, development and deployment of vehicular networks is boosted by recent advances in wireless vehicular communication techniques, such as dedicated short-range communications (DSRC), Long-Term Evolution (LTE), IEEE 802.11p and Worldwide Interoperability for Microwave Access (WiMax). Novel  protocols that can be deployed on all these communication channels and can guarantee  authentication under attacks that can be initiated from each one of these networks is an area of future research. Safeguarding one communication channel without dealing with the threats that all these networks face, will leave the IoV vulnerable to several kinds of attacks against authentication.

\subsubsection{Social Internet of Vehicles} Social Internet of Vehicles (SIoV) describes both the social interactions among vehicles \cite{277} and among drivers \cite{278}.  Ensuring authentication in the communication among vehicles cannot guarantee full protection of identities of entities if the social notion of communication is neglected \cite{69}. Future authentiation-enhancing technologies for SIoVs should be based on proven authentication-enhancing technologies for social networks and vehicular networks.

\subsection{IoE Open Issues}
Based on the definition of the Internet of Energy as an integrated dynamic network infrastructure based on standard and interoperable communication protocols that interconnect the energy network with the Internet allowing units of energy to be dispatched when and where it is needed, it is easily understood that authentication in the IoE environment is not an easy problem to solve. IoE combines M2M, V2G, IIoT (industrial Internet of things), Smart home automation, cloud services and IoS. It would be better to define IoE as an application of the IoT on the Energy domain. Authentication on the IoE domain cannot be reassured without dealing with each of the aforementioned sub domains. Security \cite{279} and hardware \cite{280} authentication techniques along with solutions dealing with middleware security \cite{281} must be combined.

\subsection{IoS Open Issues}
The major problems that the IoS networks have to face are energy efficiency and security assurance of the sensors. Intrusion Detection Systems (IDSs) and energy efficient mechanisms are not thoroughly investigated and resolved in the surveyed authentication protocols for the IoS. Raza et al. \cite{291} proposed an idea based on real-time intrusion detection for the IoT, called SVELTE. Mechanisms that can extend the SVELTE scheme for the IoS in order to be energy effiecient would be a possible research direction. Hence, future works addressing both security, mainly IDSs, and energy will have an important contribution for the authentication protocols. In addition, we believe further research is needed to develop a new framework for combining intrusion detection systems and authentication protocols for detecting and avoiding attacks in IoS.

\subsection{Pattern recognition and biometrics for the IoT}
Hybrid authentication protocols are based on two methods for identifying an individual, including, knowledge-based (e.g., the passwords) and token-based (e.g., the badges). Each method has its weakness, i.e., 1) the password can be forgotten or guessed by an adversary, and 2) the badge can be lost or stolen. Nevertheless, the safest way is the use of biometric characteristics because two people cannot possess exactly the same biometric characteristic. Hence, future works addressing pattern recognition authentication techniques along with biometrics will have an important contribution in improving authentication in the IoT. Recently new promising efforts that apply biometrics  on IoT have been proposed \cite{Shah2016328, ren2016biometrics, hossain2016toward} and the term of Internet of biometric things (IoBT) has been introduced\cite{7750970}.
Biometric technology on the other hand rises privacy and ethical issues that need to be taken in mind when designing new authentication protocols, especially for applications that deal with critical data \cite{45666}

\section{Conclusion}\label{sec:conclusion}

This paper contains a structured comprehensive overview of authentication protocols for the IoT in four environments, including, (1) Machine to machine communications (M2M), (2) Internet of Vehicles (IoV), (3) Internet of Energy (IoE), and (4) Internet of Sensors (IoS). We presented the survey articles published in the recent years for the IoT. We reviewed the major threats, countermeasures, and formal security verification techniques used by authentication protocols. We presented a side-by-side comparison in a tabular form for the current state-of-the-art of authentication protocols proposed for M2M, IoV, IoE, and IoS.

After conducting a comprehensive survey of authentication protocols, we see that the reliability of an authentication protocol depends not only on the effectiveness of the cryptography method used against attacks but also on the computation complexity and communication overhead. Therefore, in order to guarantee authentication between the machines for the IoT, we invite well-positioned researchers and practitioners to propose authentication frameworks that covers not only one but three layers, namely, the application layer, the network layer, and the sensing layer. In this paper, we also see a need for a comprehensive survey for privacy-preserving schemes for the IoT under four environments, including, M2M, IoV, IoE, and IoS. 

Authentication protocols for the IoT may be improved in terms of (1) addressing both the authentication and privacy problem, (2) developing efficient IDSs, (3) improving the computation complexity of the proposed methods, (4) improving the communication overhead of the methods, (5) developing of formal security verification techniques, (6) accounting of the  process of detecting and avoiding attacks, and (7) capturing of experts opinion in the field of computer security.

\section*{References}

\bibliography{SurveyIoT}

\begin{thebibliography}{100}
\bibitem{112}
{ISO/IEC is 9798-3}, {Entity authentication mechanisms}, part 3: Entity
  authentication using asymmetric techniques.
\end{thebibliography}

\end{document}